\documentclass{aa}  

\usepackage{graphicx}
\usepackage{txfonts}
\usepackage{natbib}
\usepackage[bookmarks=false, colorlinks=true, citecolor=blue, linkcolor=blue]{hyperref}
\def\kms{\,km\,s$^{-1}$}

\begin{document} 

\title{A search for the OH 6035\,MHz line in high-mass star-forming regions}

   \author{M. Szymczak
          \inst{1} \href{https://orcid.org/0000-0002-1482-8189}{\includegraphics[scale=0.5]{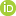}}
          \and
          P. Wolak
          \inst{1} \href{https://orcid.org/0000-0002-5413-2573}{\includegraphics[scale=0.5]{orcid.png}}
          \and
          A. Bartkiewicz
          \inst{1} \href{https://orcid.org/0000-0002-6466-117X}{\includegraphics[scale=0.5]{orcid.png}}
          \and
          M. Aramowicz
          \inst{2}
          \and
          M. Durjasz
          \inst{1}
          }

   \institute{Institute of Astronomy, Faculty of Physics, Astronomy and Informatics, Nicolaus Copernicus University, Grudziadzka 5, 87-100 Torun, Poland,
         \and
         Astronomical Institute, University of Wroc{\l}aw, ul. Kopernika 11, 51-622 Wroc{\l}aw, Poland}

  \date{Received 23 July 2020 / Accepted 2 September 2020}

% \abstract{}{}{}{}{} 
% 5 {} token are mandatory
 
  \abstract
{The excited states of OH masers detected in the environment of high-mass young stellar objects (HMYSOs) are important for improving our understanding of the physical conditions of these objects and also provide information about their magnetic fields.}
{We aim to search for excited-state OH 6035\,MHz maser emission in HMYSOs which might have escaped detection in previous surveys or were never searched for.} 
{A sample of HMYSOs derived from untargeted surveys of the 6668\,MHz methanol maser line was observed at 6035\,MHz OH transition with the Torun 32\,m radio telescope. The 6035\,MHz detections were observed in the OH 6031\,MHz line. Two-thirds of the detections were observed at least three times over a two-year period.}
{Out of 445 targets, 37 were detected at 6035\,MHz, including seven new discoveries. The 6031\,MHz line was detected towards ten 6035\,MHz sources, one of which was not previously reported. All the newly detected sources are faint with the peak flux density lower than 4\,Jy and show significant or high variability on timescales of 4 to 20\,months. Zeeman pair candidates identified in three new sources imply a magnetic field intensity of 2 to 11\,mG. Comparison of our spectra with those obtained $\sim$10\,yr ago indicates different degrees of variability but there is a general increase in the variability index on an $\sim$25\,yr timescale, usually accompanied by significant changes in the profile shape.}
{}


\titlerunning{OH 6035\,MHz line in high-mass star-forming regions}
\authorrunning{M. Szymczak et al.}

   \maketitle

\section{Introduction}\label{sec:intro}
Observations of spectral lines in the gas surrounding high-mass young stellar objects (HMYSOs) are one of the important tools to determine the physical and chemical conditions which enable the examination of mechanisms and star formation processes (\citealt{zinnecker2007}). Maser lines are of special interest in this context as useful signposts of star formation activity (e.g. \citealt{menten1991,caswell2003,breen2015}) which owing to their high levels of brightness and compactness can probe neutral gas cloudlets of a few tens of astronomical units (au) in size that reside in rotating structures such as toroids and discs (\citealt{beltran2016}) or around powerful jets (e.g. \citealt{anglada2018}).

Ground-state OH maser transitions are one of the essential signatures of HMYSOs in their early stages of formation and have been detected in numerous sites (e.g. \citealt{caswell1999,forster1999,argon2000,edris2007,beuther2019,qiao2020}). They are sometimes accompanied by the excited-state OH ($^2\Pi_{3/2}, J=5/2$) transitions at 5\,cm wavelength, where the main line 6035\,MHz dominates in most cases (\citealt{yen1969,knowles1976,smits1994,caswellvaile1995,baudry1997,caswell2001,caswell2003,avison2016}). As this excited state of OH lies immediately above the ground state, it provides a critial test for maser pumping schemes (\citealt{baudry1997,pavlakis2000,cragg2002}). OH is a  paramagnetic molecule, and therefore a significant Zeeman splitting is observed for the transitions, allowing reliable estimates of the magnetic field strength and its direction (\citealt{baudry1997,caswellvaile1995,caswell2003}). 

Most of past surveys of the excited-state OH maser transitions were commonly restricted to targets identified by ground-state OH masers (\citealt{caswellvaile1995, baudry1997}) and obviously suffer from biases. Detection of the 6.7\,GHz methanol line (\citealt{menten1991}), which is uniquely associated with star forming regions, opened a new path to identify more HMYSOs. Indeed, recent surveys of the 6.7\,GHz line resulted in detection of several previously unknown HMYSOs, enlarging the number of candidates in early stages, where the massive star is still in an active phase of accretion and is deeply embedded in the parent molecular clouds (e.g. \citealt{green2010,szymczak2012,breen2015}). Recently, the first complete untargeted survey of the  accessible southern Galactic plane for the OH 6035\,MHz line was carried out as part of the Methanol Multibeam Survey (MMB, \citealt{avison2016,avison2020}). In this paper we report the results of the OH 6035\,MHz survey of HMYSO candidates with which we aim to expand the sample of excited-state OH sources, particularly for the northern hemisphere, and to search for sources that may have escaped detection in previous observations due to variability. Observations of excited-state OH masers may allow us to find targets for multi-line maser studies with high angular resolution.

\section{Observations}\label{sect:obser}
Observations of the $^2\Pi_{3/2}, J=5/2, F=3-3$ OH transition at 6035.092\,MHz were carried out from June to September 2018 with the Torun 32\,m radio telescope. Detections were reobserved in two sessions: November-December 2018 and March-April 2019, and since then several sources have been monitored. Furthermore, all of them were also searched for the  $^2\Pi_{3/2}, J=5/2, F=2-2$ line at 6030.747\,MHz. The telescope has a half-power beam width of 6\farcm4 at these frequencies and the pointing accuracy was about 10\arcsec. The observations were pointed on the positions of 6668\,MHz methanol maser sources whose coordinates are known with sub-arcsecond accuracy in almost all cases. The sample includes all the methanol masers from the Torun catalogue \citep{szymczak2012} updated with objects above declination $-$22\degr\, from the Multibeam Methanol Survey (\citealt{green2010,breen2015}). The targets were observed in left- and right-hand circular (LHC and RHC) polarisation simultaneously using a dual-channel receiver system with the system temperature ranging from 25 to 30\,K. The IEEE convention for the handedness of polarisation was adopted and Stokes $V$ parameter was defined following the IAU convention as $V =S(\mathrm{RHC})-S(\mathrm{LHC})$, where $S(\mathrm{RHC})$ and $S(\mathrm{LHC})$ are the line flux densities for right and left circular polarisation, respectively. Spectra were obtained with an autocorrelation spectrometer in the frequency-switching mode using two banks of 4096 channels each, covering a velocity range of 95\kms\, with a velocity resolution of 0.1\kms\, after Hanning smoothing. The spectra were centred at the middle velocity of the methanol maser profiles measured relative the local standard of rest. Typical integration lasted 20\,min resulting in an rms noise level of 0.20-0.25\,Jy for a single polarisation flux density. Parameters of the receiving system were regularly measured through observations of continuum and spectral line calibrators as described in \cite{szymczak2012}. The gain of each polarisation channel was measured with a noise diode at the beginning of each ninety-second integration cycle to $\sim$10\% in absolute value and to within $\sim$4\% in relative value. The degree of circular polarisation is defined as $m_{\mathrm{C}}=V/I$, where $V$ and $I$ are Stokes parameters of circularly polarised emission and total emission, respectively.

\section{Results}\label{sec:results}
\begin{figure*}[h!]
\centering
\includegraphics[scale=1.60]{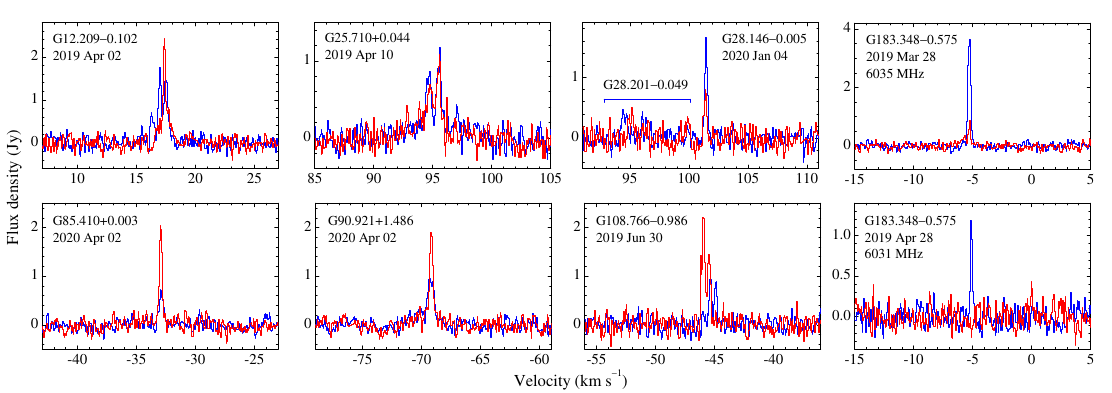}
\caption{6035\,MHz OH maser spectra of newly detected sources from the Torun observations. The new detection of the 6031\,MHz transition for one target is also shown. Blue and red lines denote LHC and RHC polarisation, respectively. Observation dates are given.}
\label{fig:new-spect}
\end{figure*}

Among 445 targets observed, the 6035\,MHz emission was detected in 37 objects of which 7 are new detections. The 6031\,MHz emission was detected towards ten 6035\,MHz objects. The parameters of spectra of new and known sources ordered by galactic longitude are listed in Tables~\ref{tab:new-det} and \ref{tab:known}, respectively. The spectra of new detections are shown in Fig.~\ref{fig:new-spect} and those of known sources are in Fig.~\ref{spec-known}. A list of non-detections is provided in Table~\ref{tab:nondetections}. 

The following notes provide information on a possible association of the 6035\,MHz emission with the 6668\,MHz masers on the basis of velocities of their maser peaks and systemic velocity of the parent molecular clouds. Estimates of the magnetic field strength of some of the masers and comments on the degree of circular polarisation are also given.

\textit{G12.209$-$00.102}. The 6035\,MHz maser emission detected in a velocity range of 16 to 18.5\kms, with $|m_{\mathrm{C}}|\approx 30-40\%$ for the strongest features, is $\sim$5\kms\, blueshifted from OH 1665\,MHz maser features \citep{argon2000} but coincides well with the 6668\,MHz methanol emission range \citep{green2010}. The 1665\,MHz emission lies within less than 2\arcsec\, from the methanol maser (\citealt{argon2000,caswell2009}) and at this position there are five methanol masers (\citealt{caswell2009, green2010}) in the telescope beam. The 6035\,MHz spectrum shows strong variability (Sect.~\ref{sec:variability}) preserving possible Zeeman splitting which corresponds to a magnetic field of +2.6 to +8.8\,mG (Table ~\ref{tab:zeeman}). 
                   
\textit{G25.709$+$00.044}. The 6035\,MHz maser shows weakly polarised emission with double peaks around 95\kms. \cite{avison2016,avison2020}
reported an almost identical 6035\,MHz spectrum towards G25.509$-$0.060 which these latter authors referred to as an isolated excited-state OH source because it has no 6668\,MHz methanol maser within $>$6\farcm5, or 22$\mu$m WISE counterpart within 1\farcm2, or 1.1\,mm ATLASGAL emission within 2\arcmin.  At this position we did not find 6035\,MHz emission with upper limit of 0.5\,Jy in 2020 April while the intensity towards G25.710+0.044 was 0.9\,Jy($>4\sigma$) at the same epoch confirming this as a new detection.

\textit{G28.146$-$00.005}. The 6035\,MHz spectrum contains a narrow feature at 101.4\kms\, which is significantly polarised ($m_{\mathrm{C}}=40\%$) and exactly coincides in velocity with the strongest feature of the 6668\,MHz maser \citep{breen2015}. In our spectrum, a side-lobe emission from G28.201$-$0.049 is seen at a velocity lower than 96.5\kms\, (\citealt{caswellvaile1995,baudry1997}). 

\textit{G85.410$+$0.003}. The 6035\,MHz spectrum is markedly polarised ($m_{\mathrm{C}}=54\%$) and the peak velocity lies at the low-velocity edge of the CH$_3$OH and H$_2$O maser spectra (\citealt{szymczak2012,urquhart2011}) very close to the systemic velocity of $-$35.8\kms\, derived from NH$_3$ lines (\citealt{urquhart2011}). This new source is associated with an embedded stellar cluster of very young stars (\citealt{persi2011}) where the 6668\,MHz maser coincides within 0\farcs3 with a compact HII region detected at centimetre wavelengths (\citealt{urquhart2009, hu2016}).

\textit{G90.921$+$1.486}. This source has 6035\,MHz polarised  emission ($m_{\mathrm{C}}=38\%$) at the same velocity as the most redshifted feature, $-$69.2\kms, of the 6668\,MHz methanol maser \citep{szymczak2012}. A tentative absorption feature centred at $-$73.5\kms\, is seen. A weak feature of the OH 1667\,MHz \citep{szymczak2000} was blueshifted by 3.3\kms\, from the 6035\,MHz feature. The methanol maser is associated with a compact HII region \citep{hu2016}.  

\textit{G108.766$-$00.986}. Double components of 6035\,MHz maser emission are seen at   the exact velocity of the strongest 6668\,MHz maser feature at $-$46.6\kms\, \citep{szymczak2012}. These are blueshifted by 6\kms\, from the systemic velocity and H$_2$O maser velocity range \citep{urquhart2011}. There is a consistent Zeeman pair seen for the whole emission implying a magnetic field of $-$10\,mG.

\textit{G183.348$-$00.575}. A highly polarised feature ($m_{\mathrm{C}}=-65\%$) is detected at a velocity of $-$5.3\kms. The strongest 6668\,MHz methanol feature is seen at nearly the same velocity \citep{szymczak2012}. The OH 6035\,MHz emission is redshifted by 4.3\kms\, from the systemic velocity \citep{wu2010}. Newly detected 6031\,MHz emission closely matches the main 6035\,MHz feature; it is completely polarised and has the narrowest profile with a width to half intensity of only 0.21\kms.

%===========================================
\begin{table*}[h!]
    \centering
    \caption{6035\,GHz OH line parameters for the new detections. The velocity range of $I$ Stokes emission ($\Delta V$), the peak velocity ($V_{\mathrm{p}}$), peak flux density ($S_{\mathrm{p}}$),  and integrated flux density ($S_{\mathrm{i}}$) for the LHC and RHC polarisation are given.}\label{tab:new-det}
    \begin{tabular}{cc@{\hskip 5pt}cc@{\hskip 5pt}c@{\hskip 5pt}c@{\hskip 5pt}c@{\hskip 5pt}c@{\hskip 5pt}c@{\hskip 3pt}c@{\hskip 5pt}c}
    \hline
         &     &    & & \multicolumn{3}{c}{LHC} && \multicolumn{3}{c}{RHC} \\
          \cline{5-7}       \cline{9-11}
    Name (l  b)    &  RA(J2000)   & Dec(J2000)  & $\Delta V$ & $V_{\mathrm{p}}$ & $S_{\mathrm{p}}$  & $S_{\mathrm{i}}$&& $V_{\mathrm{p}}$ & $S_{\mathrm{p}}$  & $S_{\mathrm{i}}$ \\
   (\degr \hspace{0.5cm} \degr)          & (h \hspace{0.2cm}  m \hspace{0.2cm}   s)   & (\degr \hspace{0.2cm} \arcmin \hspace{0.2cm} \arcsec) & (\kms) & (\kms) & (Jy)  &(Jy\kms) && (\kms) & (Jy) &  (Jy\kms) \\
    \hline
   G12.209$-$00.102   & 18 12 39.92 & $-$18 24 17.9 & 16.0;18.2 & 16.97   & 1.78 & 1.41 &&  17.35   & 2.47 & 1.14 \\
   G25.710+00.044     & 18 38 03.15 & $-$06 24 14.9 & 93.6;96.0 & 95.52   & 1.12 & 0.93 &&  95.45   & 0.85 & 0.90 \\
   G28.146$-$00.005   & 18 42 42.59 & $-$04 15 36.5 & 101.0;101.6 & 101.38   & 1.27 & 0.35 && 101.42   & 2.89 & 0.73 \\
   G85.410+00.003     & 20 54 13.68 &   +44 54 07.6 &$-$33.3;$-$32.7  & $-$32.92 & 0.74 & 0.30 && $-$32.97 & 2.00 & 0.61 \\
   G90.921+01.486     & 21 09 12.98 &   +50 01 03.6 &$-$70.5;$-$68.4  & $-$69.24 & 1.00 & 0.66 && $-$69.20 & 1.89 & 0.81 \\
   G108.766$-$00.986  & 22 58 51.18 &   +58 45 14.4 &$-$46.3;$-$44.7  & $-$45.37 & 0.95 & 0.49 && $-$45.95 & 2.21 & 1.21 \\
   G183.348$-$00.575  & 05 51 10.94 &   +25 46 17.2 & $-$6.1;$-$5.0&  $-$5.30 & 3.68 & 1.13 && $-$5.29  & 0.78 & 0.20 \\ 
   G183.348$-$00.575\tablefootmark{a}  & 05 51 10.94 & +25 46 17.2 &$-$5.4;$-$4.9 & $-$5.15 & 1.27 & 0.27 &&   & $<$0.61\tablefootmark{b} &$<$0.05  \\ 
    \hline
    \end{tabular}
\tablefoot{\tablefoottext{a}{6031\,MHz transition}, \tablefoottext{b}{$3\sigma$ level}}
\end{table*}
%============================================================

%%%%%%%%%%%%%%%%%%%%%%%%%%%%%%%%
\section{Discussion}
\subsection{Detection rate and methanol/hydroxyl luminosity ratio}
In the Galactic  longitude of 8$-$60\degr,\, our survey overlaps with the MMB untargeted observations and there are 375 methanol masers at 6.7\,GHz (\citealt{green2010,breen2015}) of which 33 (8.8\%) have an OH 6035\,MHz maser counterpart within $<10$\arcsec\, (\citealt{avison2016}). There are also ten OH sources without a methanol counterpart within 10\arcsec. In the present observation of this area the number of OH sources associated (within 10\arcsec) with the methanol maser decreased to 22 (5.9\%). The three new OH sources are not taken into account because with the beam of 6\farcm4 we were not able to discern whether or not they meet the above criterion of coincidence with the methanol masers. As a consequence of slightly lower sensitivity (by $\sim$25\%), our detection rate is lower than that inferred from the MMB survey. Similar to previous studies (\citealt{avison2016,avison2020}), we find that the OH 6035\,MHz maser emission is rather sparsely associated with the sources of the 6.7\,GHz maser line.
This implies that co-existence of 6.7\,GHz methanol and 6035\,MHz hydroxyl masers traces rather uncommon physical conditions with a narrow range of gas density of about $10^8$\,cm$^{-3}$ and low kinetic temperature of $<$50\,K (\citealt{cragg2002}). High-angular-resolution studies are required in order to decipher whether or not these transitions come from the same gas volume.

\begin{figure}[]
\includegraphics[scale=1.0]{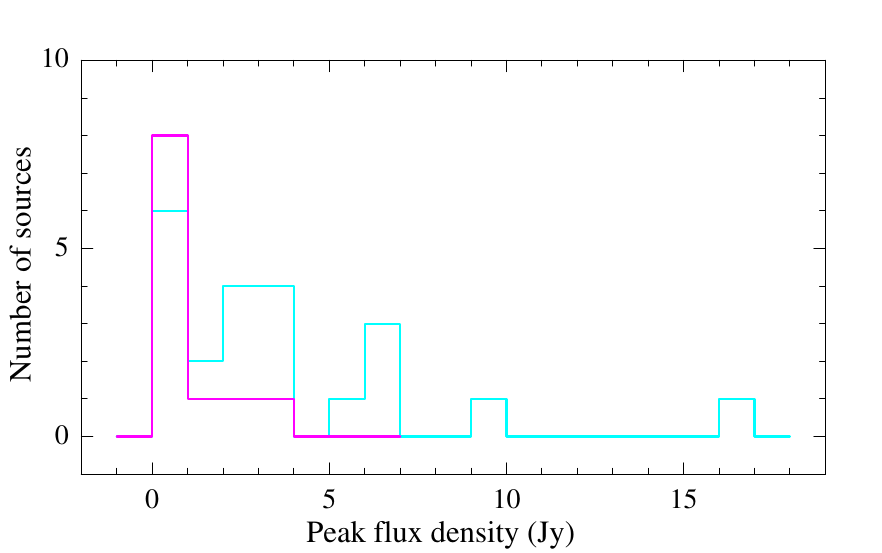}
\caption{OH 6035\,MHz peak flux of sources in the Galactic longitude range 8\degr\, to 60\degr\, from the \cite{avison2016} MMB survey. The OH sources coincide with 6.7\,GHz methanol masers within 10\arcsec. The histograms show OH detection (cyan) and non-detection (magenta) in the present survey.}
\label{fig:hist-det}
\end{figure}

Figure~\ref{fig:hist-det} shows the distribution of peak flux density of 33 OH masers at epochs (2008-2009) of MMB observations (\citealt{avison2016}) in the overlapped area. The six OH sources found by these latter authors with peak flux density above our sensitivity limit of 0.7\,Jy are not detected, while their 5 OH sources below this threshold are seen in the present survey. This implies considerable variability of 6035\,MHz transition on a timescale of $\sim$10\,yr; this point is further discussed in Sect.~\ref{sec:variability}. 

In the 8$-$60\degr\, region we found four OH masers which coincide within less than 0\farcs5 with the 6.7\,GHz sources and show distinct emission at the same velocities. Using the methanol unpublished spectra taken with the Torun 32\,m telescope (\citealt{szymczak2018}) at almost the same epochs ($\pm$3 days), we calculated the ratio of 6.7\,GHz to 6.035\,MHz isotropic luminosity for the OH velocity range. For two features of G15.034$-$00.677 centred at 21.4 and 23.5\kms\, this ratio is 1.9 and 2.6, respectively, while for the strongest OH features of G20.237+00.065 and G35.025+00.350 it is 2.5 and 3.1, respectively, and in G11.904$-$00.141 the ratio is 30.5. It is surprising that in the three objects, the luminosity of the 6.7\,GHz line is only a factor of two to three higher than that of the 6.035\,MHz line. This may suggest very specific conditions, for instance where the methanol maser is quenched due to collisions in high-density gas while the OH 6.035\,MHz maser is still excited (\citealt{cragg2002}). We stress that our estimates need to be verified with high-angular-resolution observations to confirm whether or not both transitions are really co-spacial. 

\subsection{Line width}\label{sec:linewidth}
For all spectral features with a signal-to-noise ratio (S/N) greater than five we fitted Gaussian components to estimate the line parameters. We identify 61 LHC and 66 RHC spectral features of previously known sources and 11 features at both polarisations of new detections at the 6035\,MHz transition. We also find 10 LHC and 11 RHC components at the 6031\,MHz transition for known sources and one LHC component for the new detection.
The full width at half maximum (FWHM) at 6035\,MHz ranges from 0.16 to 0.95\kms\, and the mean and median values are $0.41\pm0.05$ and 0.36\kms, respectively. These values are larger by a factor of 1.8 than those inferred from high-angular-resolution data (\citealt{desmurs1998a,fish2007}).
This discrepancy is likely due to spatial filtering out of the emission in VLBI observations where the brightest, most compact maser cloudlets
are seen and the line width is lower as compared to single dish spectra. Another possibility is that Gaussian fitting failed for blended spectra obtained with our moderate sensitivity. The mean and median values of FWHM at 6031\,MHz are $0.33\pm0.06$ and 0.36\kms\, and are consistent with those for the 6035\,MHz transition. This supports the conclusions of previous studies (e.g. \citealt{baudry1997,fish2007}). For the new detection of 6031\,MHz emission, the FWHM of the isolated component is 0.21\kms. We did not find any correlation between line width and flux density.

\subsection{Zeeman pair candidates}
We used the convention that a field directed away from the observer has a positive sign and is indicated by the RHC component at more positive velocity that the LHC component. The following coefficients were used for the magnetic field strength estimation derived from the velocity separation of each pair: $\Delta V$\kms/H(mG) = 0.056 and 0.079 at 6035\,MHz and 6031\,MHz, respectively  (\citealt{baudry1997}). 

Zeeman pairs were categorised based on a comparison of the Gaussian fits of the spectral components (Sect.\,\ref{sec:linewidth}). The fitted peak velocities and FWHM values but no peak amplitudes were used for comparison. For complex spectra, only  the prominent spectral components were considered to match pairs of components with nearby velocities and opposite senses of polarisation. We neglected pairs with one component lying on the edge of the other component.

We identified 34 LHC/RHC pairs at the 6035~MHz and 9 pairs at the 6031~MHz transition; see Table~\ref{tab:zeeman}, where we also list the fitted peak velocities and flux densities,  demagnetized velocities, field strength, reliability of Zeeman pair identification, and field strength from the literature. The inferred magnetic field ranges from 0.1 to 12.0\,mG as indicated by the splitting of the 6035\,MHz line and from 0.2 to 9.0\,mG of the 6031\,MHz line. The mean and median values are 4.6 and 4.4\,mG (6035\,MHz), and 4.6 and 3.9\,mG (6031\,MHz).

The possible Zeeman pairs listed in Table\,\ref{tab:zeeman} should be treated with care, especially those labelled  `B'. All the pairs need to be verified with high-angular-resolution observations to definitively demonstrate that the candidate Zeeman pairs spatially coincide and are thus genuinely associated. Nevertheless, our estimates of the magnetic field seem to be consistent with those reported in the literature (Table\,\ref{tab:zeeman}); in most cases, we obtained similar field strength and in all cases the same field direction. 

\citet{caswell1997} pointed out that there was no field greater than about 10~mG. Therefore, it will be important to verify, using interferometric data, the magnetic field strengths in G12.681$-$00.182, G69.540$-$00.976, and G108.7666$-$00.986, where our estimated values exceeded 10~mG. Moreover, in G15.035$-$00.677, G45.467$+$00.053, and G80.861$+$00.383 the field reversal is seen and follow-up studies would be desirable to uncover the magnetic field morphology. 

Three objects (G11.034+00.062, G11.904$-$00.141 and G15.035$-$00.677) were observed in full polarisation by \cite{green2015}. For the first two sources, our estimates of the magnetic field strength are only roughly (35\%) consistent with the values of these latter authors (Table~\ref{tab:zeeman}). This could be due to the fact that our spectra are much noisier than theirs. In the case of the bright source G15.035$-$00.677 whose 6031\,MHz profile was identified as a Zeeman triplet candidate (\citealt{green2015}), we obtained field strengths similar to those reported in the literature (Table~\ref{tab:zeeman}).

\subsection{Variability}\label{sec:variability}
Most of our newly detected sources were observed three or more times over a period of less than two years. To quantify their variability we used the variability index given by
%%%%%%%%%%%%%%%
\begin{equation}
 vi = {(S_{\rm max} - \sigma_{\rm max}) - (S_{\rm min} + \sigma_{\rm min})\over (S_{\rm max} - \sigma_{\rm max}) + (S_{\rm min} + \sigma_{\rm min})}
\end{equation}
%%%%%%%%%%%%%%%
\noindent
which is a measure of the amplitude of the variability of the spectral feature. Here, $S_{\rm max}$ and $S_{\rm min}$ are the highest and lowest measured flux densities, respectively, and $\sigma_{\rm max}$ and $\sigma_{\rm min}$ are the uncertainties in these measurements. The variability index for the strongest features on a timescale of nearly two years, $vi_{2}$, is listed in Table~\ref{tab:var-index}. We restrict our analysis to sources with a peak flux density greater than 1\,Jy to eliminate spurious effects. For previously known objects, we estimated the variability indices relative to the 2008-2009 measurements of \cite{avison2016} $(vi_{10}$) and the 1993-1994 observations by \cite{caswellvaile1995} ($vi_{25}$). These variability indices are added to Table~\ref{tab:var-index} and for sources with complex spectra the velocity and polarisation of the considered feature are given in the notes. 

\begin{table}
\centering
\caption{Variability indices of 6035\,MHz emission for the main feature on timescales of 2 ($vi_2$), 10 ($vi_{10}$), and 25 ($vi_{25}$) years. For sources with complex spectra, the velocity and polarisation of the feature used are noted. New detections are in bold.}\label{tab:var-index}
\begin{tabular}{lccc}
\hline
    Source & $vi_2$ & $vi_{10}$ & $vi_{25}$  \\
    \hline
    G10.320$-$00.259  &  0.04  &  0.02  &         \\
%    G10.627$-$00.384  &  0.11  &  0.10  &  0.28   \\
    G10.958+00.022    &  0.02  &  0.12  &         \\
    G11.034+00.062    &  0.04  &  0.03  &  0.79   \\
    G11.904$-$00.141  &  0.30  &  0.78  &  0.78   \\
{\bf    G12.209$-$00.102}  &  0.72  &        &         \\
    G12.681$-$00.182  &  0.42  &  0.73  &         \\
    G15.035$-$00.677$^{(1),(7)}$  &  0.06  &  0.40  &  0.48  \\
 %   G19.486+00.151    &  0.06  &  0.63  &  0.78   \\
    G20.237+00.065    &  0.30  &  0.24  &  0.35   \\
    G24.148$-$00.009  &  0.34  &  0.54  &         \\
    G25.650+01.049    &  0.03  &  0.09  &         \\
{\bf    G25.710+00.044}    &  0.21  &        &         \\
{\bf    G28.146$-$00.005}  &  0.25  &        &         \\
    G28.201$-$00.049  &        &  0.31  &  0.34   \\
    G30.771$-$00.804  &        &  0.08  &         \\
    G34.257+00.153$^{(2),(7)}$&        &  0.46  &  0.70  \\
    G34.267$-$00.210  &  0.02  &        &         \\
 %   G34.284+00.184  &          &        &      &   \\
    G35.025+00.350    &  0.18  &  0.26  &  0.44   \\
 %   G40.425+00.700    &        &  0.29  &  0.71   \\
    G43.149+00.013    &  0.12  &  0.16  &  0.44   \\
%    G43.796$-$00.127  &  0.21  &  0.19  &  0.20   \\
    G45.467+00.053$^{(3)}$&  0.02  &  0.10  &  0.20  \\
    G48.990$-$00.299  &  0.16  &  0.20  &         \\
    G49.490$-$00.388$^{(4),(7)}$  &  0.02  &  0.24  &  0.67  \\
%    G69.540$-$00.976  &          &        &      &   \\
    G80.861+00.383    &  0.39  &        &         \\
    G81.871+00.781$^{(5)}$  &  0.18    &        &     \\
{\bf    G85.410+00.003}  &  0.17   &        &         \\
%    G90.921+01.486  &          &        &      &   \\
%    G98.036+01.446  &  0.10    &        &        \\
{\bf    G108.766$-$00.986} &   0.22  &        &       \\
    G111.542+00.777   &  0.28   &        &       \\
    G133.947+01.064$^{(6)}$   & 0.11        &        &    \\
{\bf    G183.348$-$00.575}  &   0.53  &        &      \\
 %   G189.030+00.784  &          &        &      &   \\
%    G208.997$-$19.387  &          &        &      &   \\
    \hline
    \end{tabular}
\tablefoot{$^{(1)}$21.5\kms\, LHC; $^{(2)}$58.4\kms\, RHC; $^{(3)}$66.38\kms\, LHC; $^{(4)}$55.0\kms\, RHC; $^{(5)}$7.87\kms\, RHC; $^{(6)}$$-$43.3\kms\, LHC; $^{(7)}$large changes in the spectrum shape}
\end{table}

Of 25 sources with reliably estimated $vi_2$, 8 show little or no variation within the noise, 12 show moderate or significant variability (0.1$<vi_2<$0.3), and 5 sources show large variability, that is, a factor of $\ge$2 ($vi_2>$0.3). We note that all the newly detected sources show significant or high variability. We suggest that the considerable variability of G12.209$-$00.102, G25.710+0.044, and G28.146$-$0.005 is the cause of the non-detection in the more sensitive MMB survey (\citealt{avison2016}).

\begin{figure}[]
\includegraphics[scale=1.25]{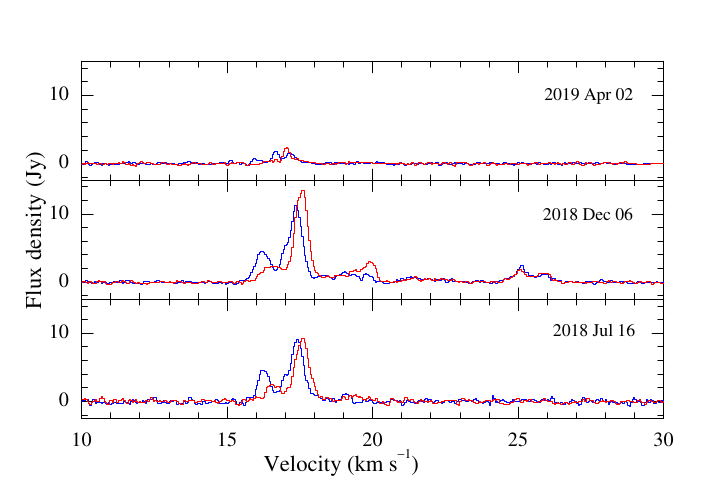}
\caption{OH 6035\,MHz spectra of G12.209$-$00.102 at three epochs. Blue and red lines denote LHC and RHC polarisation, respectively.}
\label{fig:g12-209var}
\end{figure}

The variability index of the main feature does not give the complete picture of variability for sources with complex spectra. Figure~\ref{fig:g12-209var} shows the spectra of G12.209$-$00.102 at three epochs spanning almost 9\,months. In July and December 2018, the shape of the spectrum in the velocity range of 15.5 to 20.2\kms\, was preserved but it was completely rebuilt in April 2019 when the intensity decreased by a factor of seven and polarisation properties changed remarkably. Clearly detected emission at velocities higher than 24.7\kms\, with a peak flux of 2.7\,Jy appeared only in December 2018. This case proves that remarkable variations of OH 6035\,MHz emission occur on timescales of less than 4\,months. No significant variations in a 6\,month interval were reported by \cite{caswellvaile1995}.   

\begin{figure}[]
\includegraphics[scale=1.0]{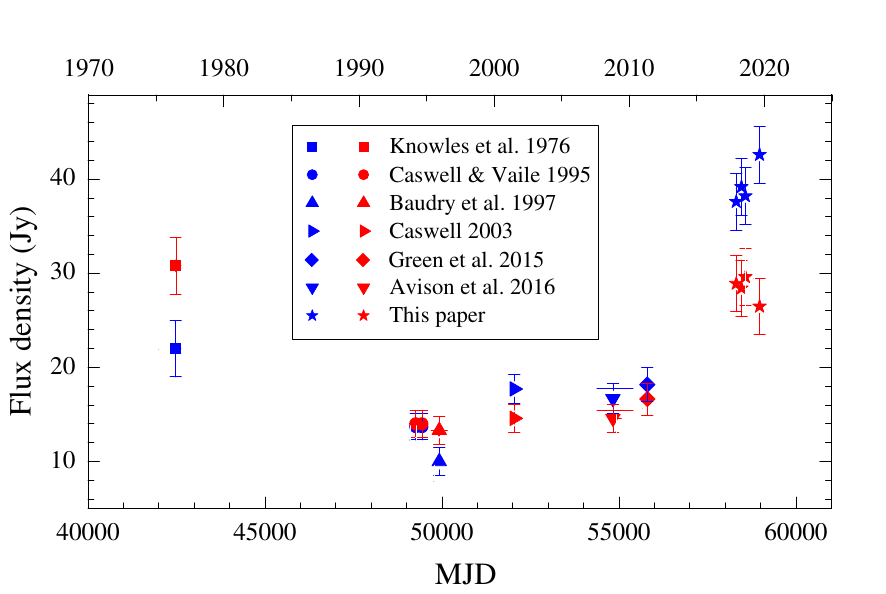}
\caption{Long term variability of the main feature (21.5\kms) of G015.035$-$00.677 at 6035\,MHz. Blue and red symbols denote LHC and RHC polarisation, respectively.}
\label{fig:g15-var}
\end{figure}

A general trend can be seen in terms of an increase in variability index with increasing timescales (Table~\ref{tab:var-index}). Figure~\ref{fig:g15-var} shows the light curve of the main feature in G15.035$-$00.677. The first published spectrum with the flux density scale was obtained in 1975 (\citealt{knowles1976}) and contains two spectral features; the most prominent of the two corresponds to our 21.45\kms\ feature and the emission in RHC polarisation was stronger than that in LHC polarisation. The RHC flux density  of this latter feature decreased by a factor of 2.2 after 18.6\,yr (\citealt{caswellvaile1995}), and then remained stable within 15\% over a period of 18\,yr and increased by a factor of two after 10\,yr. The LHC flux density followed the same course. During a low plateau state the difference between polarisation diminished and then the LHC signal became dominant. The flux ratio of the 21.45 and 22.65\kms\, features varied from 1.8 (\citealt{knowles1976}) to 0.96 (\citealt{caswell2003}) and then to 2.1 during our observations. The case of G15.035$-$00.677 depicts significant variations in the intensity, spectrum shape, and polarisation over a period of 45\,yr. We note significant changes on a similar timescale for the 6035 and 6031\,MHz masers of G133.947+01.064 (W3OH) when comparing our spectra with those reported in the literature (e.g. \citealt{moran1978,desmurs1998,fish2007}).

\subsection{Unusual flux ratio in G49.490$-$00.388}

\begin{figure}[]
\includegraphics[scale=1.25]{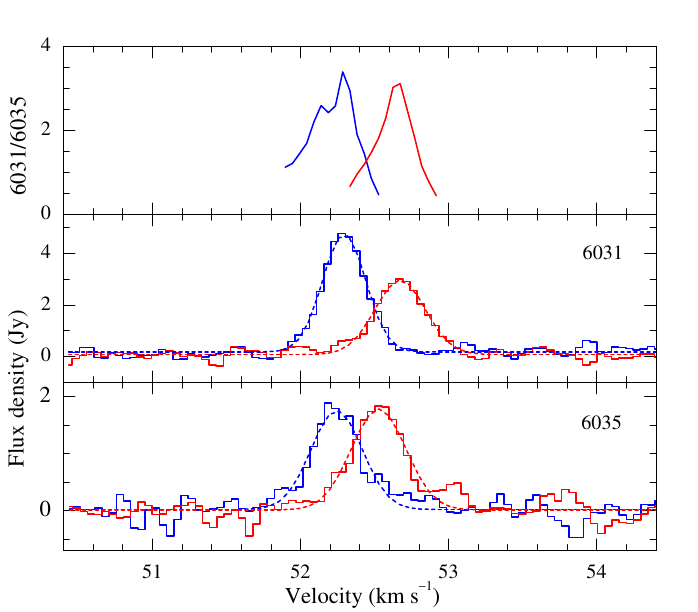}
\caption{Part of OH spectra in G49.490$-$00.388 showing a rare case where the maser intensity at 6031\,MHz surpasses that at 6035\,MHz. The upper panel shows the line ratio for each polarisation. Blue and red lines are LHC and RHC polarisation, respectively. The dashed lines correspond to the Gaussian fits.}
\label{fig:g49-both}
\end{figure}

Towards the G49.490$-$00.388 site, the OH emission near the velocity of 52.4\kms\, shows striking characteristics; the 6031/6035\,MHz flux density ratio exceeds unity for almost the total width of the profile, reaching a peak value of 3.4 and 3.1 for LHC and RHC polarisation, respectively (Fig.\,\ref{fig:g49-both}). This phenomenon in the source was discussed by \cite{caswell2003} who also noted similar properties in G345.010+01.792 and G353.410$-$00.360 observed by \cite{smits1994}, but in the first of them this feature was short lived (\citealt{caswell2003}). \cite{baudry1997} noted the well-known source G109.871+2.114 (Cep\,A) as the exceptional case with the 6031/6035 flux ratio of order unity. This source probably shows significant variability and was not detected in the present survey.

In G49.490$-$00.388, the line profiles of both transitions match in velocity and polarisation. A magnetic field strength derived from the velocity separation of peaks in each sense of circular polarisation is +4.8$\pm$0.3 and +5.2$\pm$0.4\,mG for 6031 and 6035\,MHz, respectively, and is consistent with that observed $\sim$18\,yr ago but at a velocity near 53.2\kms\, (\citealt{caswell2003}). \cite{avison2016} reported the MMB observations taken in 2008-2009 where 6031\,MHz RHC polarised emission at 52.7\kms\, is stronger by a factor of 1.3 than the counterpart emission at 6035\,MHz. The data from 1994 suggest a similar flux ratio (\citealt{baudry1997,desmurs1998}). In turn, the spectra from 1973 imply a ratio of 0.6 and 0.4 for the LHC and RHC polarisation, respectively, when the profiles were seen near 53.2\kms\, (\citealt{rickard1975}). The magnetic field strength estimates at that epoch of +4.1 to +5.7\,mG are fully consistent with ours. We conclude that the flux ratio of the feature varies considerably, exceeding unity for a period of at least 25\,yr. This emission appeared at slightly different velocities and the spectra taken at four epochs: 1994 (\citealt{baudry1997, desmurs1998}), 2001 (\citealt{caswell2003}), 2008-2009 (\citealt{avison2016}), and 2019 (this paper) suggest a velocity drift of 0.03\kms\,yr$^{-1}$. 

High-angular-resolution observations of a few sources revealed a good match of positions between the 6031 and 6035\,MHz maser components (\citealt{desmurs1998,desmurs1998a,etoka2005}) suggesting co-propagation of both lines. According to model calculations, these transitions probe the gas for low gas temperature ($<$70\,K), high density ($3\times10^{7-8}$\,cm$^{-3}$), and OH column densities greater than $2\times10^{17}$\,cm$^{-2}$ (\citealt{cragg2002}). In the models of these latter authors, the 6031\,MHz line closely accompanies the 6035\,MHz line but is usually weaker. This prediction is consistent with observations for which the typical value of the peak flux ratio of 6031/6035 ranges from 0.14 to 0.5 (\citealt{baudry1997,caswell2003,avison2016}). Thus, the excess of 6031\,MHz emission in G49.490$-$00.388 could be produced under some special conditions that could be extremely rare or transient (\citealt{caswell2003}) and that have not been explored in models. As the ratio increased over more than four decades and the profile slightly drifts in velocity, the maser may emerge from a region accelerated by stellar wind or outflow where physical conditions are readily deviating from typical parameters tested in the maser models (\citealt{cragg2002}).

%%%%%%%%%%%%%%%%%%%%%%%%%%
\section{Conclusions}
The detection rate of excited-state OH transition of 6\% implies a rare association of OH 6035\,MHz and CH$_3$OH 6668\,MHz masers. Nevertheless, we identified three objects with possible co-propagation of both transitions. For these sources, the ratio of CH$_3$OH/OH isotropic luminosities is only between two and three and may indicate the gas cloudlets with a narrow range of gas density of about $10^8$\,cm$^{-3}$ and low kinetic temperature of $<$50\,K. This possibility needs to be examined with high-angular-resolution observations.    

All the newly detected OH maser sources show variability greater than a factor of 1.4 to 6.1 on timescales of 4-20\,months while their non-detection in previous more sensitive surveys suggests extraordinary variations on timescales of several years. For previously known sources, we saw a general trend of an increase in variability index for longer (10-25\,yr) timescales.  

A rare case of a source with the maser intensity at 6031\,MHz surpassing that at 6035\,MHz was confirmed. Inspection of the available data revealed one feature with considerable variations of the 6031/6035 flux ratio, exceeding unity for a period of 25\,yr. This phenomenon cannot be explained by the standard models but observational characteristics such as a drift in velocity suggest that it occurs in a region accelerated by stellar wind or outflow. Further monitoring and interferometric studies are required to understand this unusual case.  

\begin{acknowledgements}
The 32\,m radio telescope is operated by the Institute of Astronomy, Nicolaus Copernicus University and supported by the Polish Ministry of Science and Higher Education SpUB grant. We thank the staff and students for assistance with the observations. This research has made
use of the SIMBAD data
base, operated at CDS (Strasbourg, France) and NASA’s Astrophysics Data System Bibliographic Services. We acknowledge support from the Polish  National Science Centre grant 2016/21/B/ST9/01455.
\end{acknowledgements}

\bibliography{librarian}{}
\bibliographystyle{aa}

\begin{appendix}
\section{Supplementary materials}

\begin{figure*}
\centering
\resizebox{0.99\hsize}{!}{\includegraphics[scale=0.4]{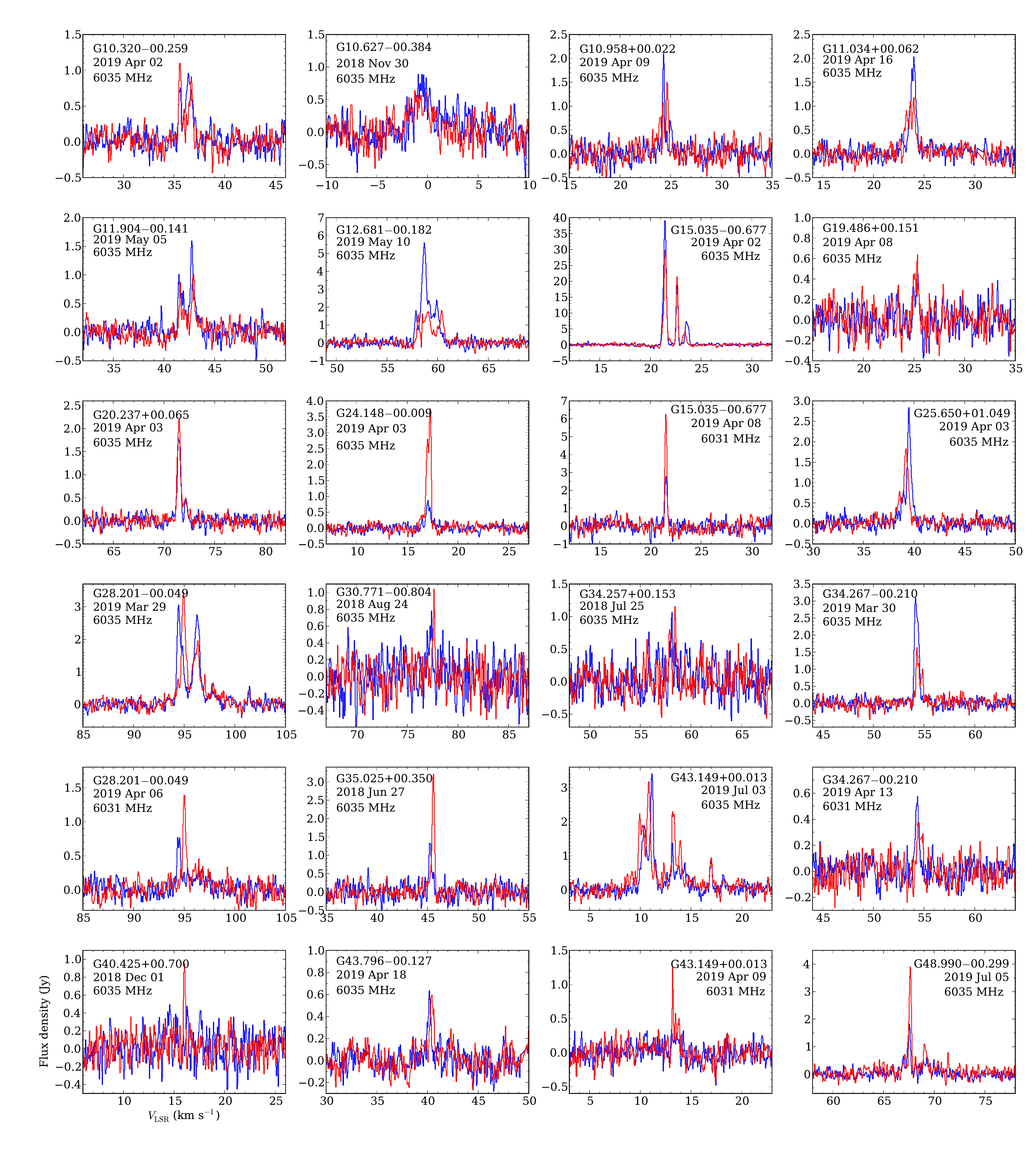}}
\caption{\label{spec-known} 6035\,MHz and 6031\,MHz OH maser spectra as detected in the Torun survey. Red and blue lines are RHC and LHC polarisations, respectively.}
\end{figure*}

\addtocounter{figure}{-1}
\begin{figure*}
\centering
\resizebox{0.99\hsize}{!}{\includegraphics[scale=0.4]{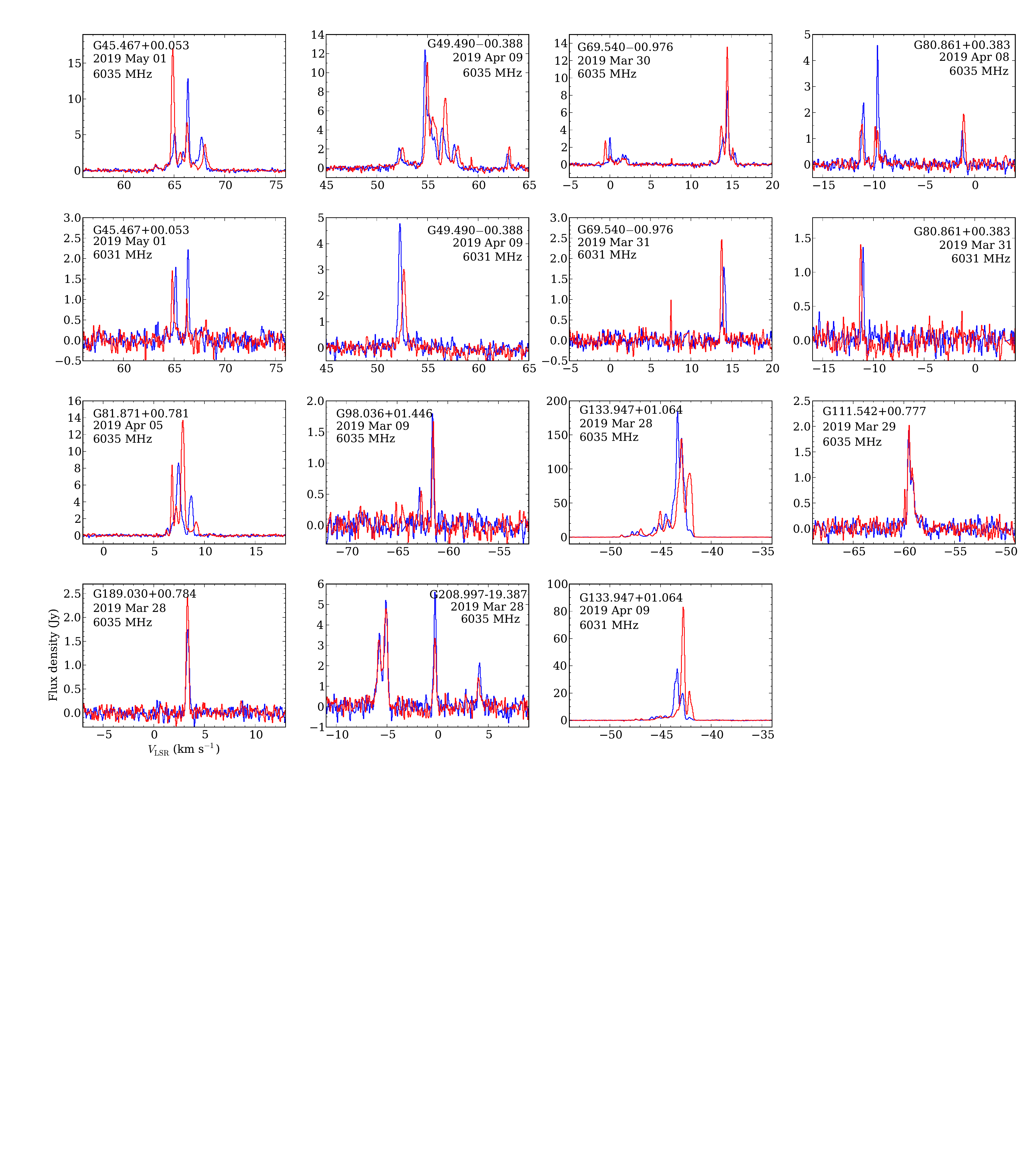}}
\caption{continued}
\end{figure*}

\begin{table*}[p]
    \centering
    \caption{Properties of 6035 and 6031\,MHz OH maser emission for previously known sources. The following parameters are listed; velocity range of $I$ Stokes emission ($\Delta V$), peak velocity ($V_{\mathrm{p}}$), peak flux density ($S_{\mathrm{p}}$), and integrated flux density ($S_{\mathrm{i}}$) for the LHC and RHC polarisations.}
    \begin{tabular}{cc@{\hskip 5pt}cc@{\hskip 5pt}c@{\hskip 5pt}c@{\hskip 5pt}c@{\hskip 5pt}c@{\hskip 5pt}c@{\hskip 3pt}c@{\hskip 5pt}c}
    \hline
         &     &    & & \multicolumn{3}{c}{LHC} && \multicolumn{3}{c}{RHC} \\
          \cline{5-7}       \cline{9-11}
    Name (l  b)    &  RA(J2000)   & Dec(J2000)  & $\Delta V$ & $V_{\mathrm{p}}$ & $S_{\mathrm{p}}$  & $S_{\mathrm{i}}$&& $V_{\mathrm{p}}$ & $S_{\mathrm{p}}$  & $S_{\mathrm{i}}$ \\
   (\degr \hspace{0.5cm} \degr)          & (h \hspace{0.2cm}  m \hspace{0.2cm}   s)   & (\degr \hspace{0.2cm} \arcmin \hspace{0.2cm} \arcsec) & (\kms) & (\kms) & (Jy)  &(Jy\kms) && (\kms) & (Jy) &  (Jy\kms) \\
    \hline 
    G10.320--00.259  & 18 09 23.30 & $-$20 08 06.90  & 35.3;37.2 & 36.41 & 0.96 & 0.97 & & 35.59 & 1.10 & 0.69  \\
    G10.627--00.384 & 18 10 29.22 & $-$19 55 41.10 &  $-$2.2;1.1 & $-$0.91 & 0.89 & 2.02 & & $-$1.24 & 0.64 & 1.20 \\ 
    G10.958$+$00.022  & 18 09 39.32 & $-$19 26 28.00 & 23.8;25.3 & 24.32 & 2.10 & 0.77 & & 24.66 & 1.50 & 0.70 \\
    G11.034$+$00.062 & 18 09 39.84 & $-$19 21 20.30 & 23.0;24.5 & 23.99 & 2.03 & 1.27 & & 23.99 & 1.18 &  0.94 \\ 
    G11.904--00.141 & 18 12 11.44 & $-$18 41 28.60 & 41.3;43.7 & 42.74 & 1.60 & 1.28 & & 42.94 & 1.01 & 0.88\\
    G12.681--00.182 & 18 13 54.75 & $-$18 01 46.60 & 57.6;60.8 & 58.69 & 5.59 & 5.83 & & 60.39 & 1.82 & 2.91\\
    G15.035--00.677 & 18 20 24.78 & $-$16 11 34.60 & 21.0;23.9& 21.42 & 39.04 & 20.48 & & 21.46 & 29.81 & 16.34\\
    & 6031 & & 21.3;21.8 & 21.51 & 6.25 & 1.49 & & 21.56 & 2.78 & 0.74\\
    G19.486$+$00.151 & 18 26 00.39 & $-$11 52 22.60 & 24.7,25.7 & 25.40 & 0.45 & 0.19 & & 25.35 & 0.64 & 0.25\\
    G20.237$+$00.065 & 18 27 44.56 & $-$11 14 54.30  & 71.07;72.5 &71.42 & 1.79 & 0.83 & & 71.51 & 2.20 & 1.03 \\
    G24.148--00.009 & 18 35 20.94 & $-$07 48 55.67 & 16.3;17.7 & 17.06 & 0.87 & 0.35 & & 17.25 & 3.75 & 1.73 \\ 
    G25.650$+$01.049 & 18 34 20.90 & $-$05 59 42.20 & 38.1;40.2 & 39.52 & 2.84 & 0.16 & & 39.23 & 1.83 & 1.28 \\
   G28.201$-$00.049 & 18 42 58.08 & $-$04 13 56.20 &  94.0;98.9 & 94.46 & 3.05 & 4.20 & & 94.94 & 3.45 & 4.04 \\
   & 6031 &  &  94.3;97.6 &  94.30 & 0.77 & 0.76 & & 95.04 & 1.39 & 0.95 \\
    G30.771$-$00.804  & 18 50 21.55  & $-$02 17 24.00 & 76.8,77.9 & 77.41 & 0.78 & 0.27 & & 77.65 & 1.04 & 0.20\\ 
    G34.257$+$00.153 & 18 53 18.63 & $+$01 14 57.40 & 57.5;58.7 & 58.14 & 1.06 & 0.25 & & 58.43 & 1.15 & 0.50\\ 
    G34.267$-$00.210 & 18 54 37.25 & $+$01 05 33.70 & 53.9;55.0 & 54.16 & 3.09 & 1.47 & & 54.31 & 1.64 & 0.82 \\ 
    & 6031 && 53.9;54.8 & 54.35 & 0.58 & 0.23 & & 54.40 & 0.38 & 0.16 \\
    G35.025$+$00.350 & 18 54 00.66 & $+$02 01 19.30 & 45.1;45.7 & 45.21 & 1.33 &  0.45 & &  45.60 & 3.21 & 0.85 \\
    G40.425$+$00.700 & 19 02 39.62 & 06 59 10.50 & 15.8;16.1 & $-$ & $-$ & 0.05 & & 16.03 & 0.95 & 0.15 \\
    G43.149$+$00.013 & 19 10 11.05 & $+$09 05 20.40 & 8.5;18.6 & 11.15 & 3.41 & 3.95 & & 10.86 & 3.17 & 5.67 \\
    & 6031 && 12.9;14.0 &$-$ & $-$ & 0.16 & &  13.19 & 1.25 & 0.37 \\ 
    G43.796--00.127 & 19 11 53.97 & $+$09 35 53.50 &  39.8;40.7 & 40.18 & 0.63 & 0.22 & & 40.43 & 0.60 & 0.35\\ 
    G45.467$+$00.053 & 19 14 24.15 & $+$11 09 43.00 & 62.7;68.6 & 66.38 & 12.82 &  10.63 & & 64.87 & 17.06 & 12.16\\
    & 6031 && 63.9;68.3 & 66.38 & 2.22 & 1.03 && 64.83 & 1.70 & 0.94\\
    G48.990--00.299 & 19 22 26.13 & $+$14 06 39.78 & 66.7;69.6 & 67.55 & 1.81 & 1.34 & & 67.65 & 3.89 & 1.96\\
    G49.490--00.388 & 19 23 43.95 & $+$14 30 34.20 & 52.1;63.3 &  54.76 & 12.39 & 13.25 & &55.00 & 11.07 & 14.80\\ 
     & 6031 & &  52.1;53.1 & 52.28 & 4.77 & 1.93 & &  52.67 & 3.01 & 1.40 \\
    G69.540--00.976 & 20 10 09.07 & $+$31 31 34.86&  $-$0.9;15.7 & 14.53 & 8.52 & 7.53 & & 14.48 & 13.56 & 8.69 \\
     & 6031 & & 13.4;14.3 & 14.05 & 1.80 & 0.69 && 13.81 & 2.46 & 0.73\\
    G80.861$+$00.383 & 20 37 00.96 & $+$41 34 55.70 &  $-$11.5;$-$0.8 & $-$9.60 & 4.58 & 2.45 &&  $-$1.10 & 1.94 & 1.70 \\
    & 6031 & & $-$11.2;$-$10.7 & $-$11.01 & 1.37 & 0.28 && $-$11.26 & 1.40 &  0.19 \\ 
    G81.871$+$00.781 & 20 38 36.42 & $+$42 37 34.56 &  6.1;9.5 & 7.44 & 8.63 & 7.09 && 7.87 & 13.68 & 9.34 \\ 
    G98.036$+$01.446 & 21 43 01.43   & $+$54 56 17.75 &  $-$63.0;$-$61.3 & $-$62.76 & 0.61 & 0.49 && $-$62.60 & 0.55 & 0.49\\ 
    G111.542$+$00.777 & 23 13 45.36   & $+$61 28 10.55 & $-$60.4;$-$58.1 & $-$59.52 & 1.91 & 0.92 & & $-$59.47 & 2.02 & 1.18 \\ 
    G133.947$+$01.064 & 02 27 03.82 & $+$61 02 25.4 & $-$49.1;$-$41.1 & $-$43.31 & 184.16 &  190.1 & & $-$42.92 & 144.04 & 181.9\\
     & 6031 & & $-$47.7;$-$41.4 & $-$43.35 & 37.72 & 30.84 & & $-$42.77 & 83.04 & 45.85 \\ 
    G189.030$+$00.784 & 06 08 40.67 & $+$21 31 06.90 & 3.0;3.7 & 3.37 & 1.74 & 0.57 & & 3.32 & 2.43 & 0.70\\
    G208.997--19.387 & 05 35 14.50 & $-$05 22 45.00 & $-$6.4;4.6 & $-$0.27 & 5.60 &5.92 & & $-$5.07 & 4.80 & 4.42 \\
    \hline 
    \end{tabular}
    \label{tab:known}
\end{table*}

\begin{table*}
\caption{List of non-detections. The central velocity $V_{\mathrm{c}}$ of observation is given. Previously known sources are bolded and reference for detection is given in the last column. \label{tab:nondetections}}
\centering
\begin{tabular}[c]{c c c c c}
\hline
\hspace{\fill} Name (l    b) \hspace{\fill} & \hspace{\fill} RA (J200) \hspace{\fill} & \hspace{\fill} Dec (J2000) \hspace{\fill} & \hspace{\fill} V$_c$(6668)  \hspace{\fill} & Ref. \\
 (\degr \hspace{0.5cm} \degr)          & (h \hspace{0.2cm}  m \hspace{0.2cm}   s)   & (\degr \hspace{0.2cm} \arcmin \hspace{0.2cm} \arcsec) & (\kms) & \\
\hline
G08.317$-$00.096 & 18 04 36.02  &21 48 19.60  & 46.9 \\ 
{\bf G08.669$-$00.356} & 18 06 18.99  & $-$21 37 32.20  & 39.2 & 1 \\  
G08.683$-$00.368 & 18 06 23.49  & $-$21 37 10.20  & 43.1 \\ 
G08.832$-$00.028 & 18 05 25.66  & $-$21 19 25.50  & $-$3.9\\ 
G08.872$-$00.493 & 18 07 15.32  & $-$21 30 54.40  & 23.2 \\ 
G09.215$-$00.202 & 18 06 52.84  & $-$21 04 27.50  & 45.5 \\ 
{\bf G09.621$+$00.196 } & 18 06 14.67  & $-$20 31 32.40  & 1.2  & 2 \\ 
{\bf G09.619$+$00.193} & 18 06 14.92  & $-$20 31 44.30  & 5.5 & 2 \\ 
G09.986$-$00.028 & 18 07 50.12  & $-$20 18 56.50  & 42.2 \\ 
G10.205$-$00.345 & 18 09 28.43  & $-$20 16 42.50  & 6.6 \\ 
\\
G10.287$-$00.125 & 18 08 49.36  & $-$20 05 59.00  & 4.6 \\ 
G10.299$-$00.146 & 18 08 55.54  & $-$20 05 57.50  & 19.9 \\ 
G10.323$-$00.160 & 18 09 01.46  & $-$20 05 07.80  & 11.5 \\ 
G10.342$-$00.142 & 18 08 59.99  & $-$20 03 35.40  & 15.4 \\ 
G10.356$-$00.148 & 18 09 03.07  & $-$20 03 02.20  & 49.9 \\ 
G10.444$-$00.018 & 18 08 44.88  & $-$19 54 38.20  & 73.3 \\ 
G10.472$+$00.027 & 18 08 38.20  & $-$19 51 50.10  & 75.0 \\ 
G10.480$+$00.033 & 18 08 37.88  & $-$19 51 16.10  & 59.5 \\ 
G10.629$-$00.333 & 18 10 17.98  & $-$19 54 04.80  & $-$8.1 \\ 
G10.724$-$00.334 & 18 10 30.03  & $-$19 49 06.80  & -2.2 \\ 
\\
G10.822$-$00.103 & 18 09 50.52  & $-$19 37 14.10  & 72.0 \\ 
G10.886$+$00.123 & 18 09 07.98  & $-$19 27 21.80  & 17.1 \\ 
G11.109$-$00.114 & 18 10 28.25  & $-$19 22 29.10  & 23.9 \\ 
G11.497$-$01.485 & 18 16 22.13  & $-$19 41 27.10  & 6.6 \\ 
G11.903$-$00.102 & 18 12 02.70  & $-$18 40 24.70  & 33.9 \\ 
G11.936$-$00.150 & 18 12 17.29  & $-$18 40 02.60  & 48.5 \\ 
G11.936$-$00.616 & 18 14 00.89  & $-$18 53 26.60  & 32.2 \\ 
G11.992$-$00.272 & 18 12 51.19  & $-$18 40 38.40  & 59.8 \\ 
G12.025$-$00.031 & 18 12 01.86  & $-$18 31 55.70  & 108.2 \\ 
G12.112$-$00.126 & 18 12 33.39  & $-$18 30 07.60  & 39.9 \\ 
\\
G12.181$-$00.123 & 18 12 41.00  & $-$18 26 21.90  & 29.7 \\ 
G12.199$-$00.034 & 18 12 23.44  & $-$18 22 50.90  & 49.3 \\ 
G12.202$-$00.120 & 18 12 42.93  & $-$18 25 11.80  & 26.4 \\ 
G12.203$-$00.107 & 18 12 40.24  & $-$18 24 47.50  & 20.5 \\ 
G12.265$-$00.051 & 18 12 35.40  & $-$18 19 52.30  & 68.3 \\ 
G12.526$+$00.016 & 18 12 52.04  & $-$18 04 13.60  & 42.6 \\ 
G12.625$-$00.017 & 18 13 11.30  & $-$17 59 57.60  & 21.5 \\ 
G12.776$+$00.128 & 18 12 57.57  & $-$17 47 49.20  & 32.8 \\ 
G12.889$+$00.489 & 18 11 51.40  & $-$17 31 29.60  & 39.2 \\ 
G12.904$-$00.031 & 18 13 48.27  & $-$17 45 38.80  & 58.8 \\ 
\\
G12.909$-$00.260 & 18 14 39.53  & $-$17 52 00.00  & 39.8 \\ 
G13.179$+$00.061 & 18 14 00.96  & $-$17 28 32.50  & 46.4 \\ 
G13.657$-$00.599 & 18 17 24.27  & $-$17 22 12.50  & 51.1 \\ 
G13.696$-$00.156 & 18 15 51.05  & $-$17 07 29.60  & 99.3 \\ 
G13.713$-$00.083 & 18 15 36.99  & $-$17 04 31.80  & 43.5 \\ 
G14.101$+$00.087 & 18 15 45.80  & $-$16 39 09.70  & 15.3 \\ 
G14.230$-$00.509 & 18 18 12.59  & $-$16 49 22.80  & 25.3 \\ 
G14.331$-$00.641 & 18 18 53.80  & $-$16 47 46.60  & 22.6 \\ 
G14.390$-$00.020 & 18 16 43.77  & $-$16 27 01.00  & 26.9 \\ 
G14.457$-$00.143 & 18 17 18.79  & $-$16 27 57.50  & 43.2 \\ 
\\
G14.490$+$00.014 & 18 16 48.06  & $-$16 20 45.00  & 20.2 \\ 
G14.521$+$00.155 & 18 16 20.73  & $-$16 15 05.50  & 4.1 \\ 
G14.604$+$00.017 & 18 17 01.14  & $-$16 14 38.00  & 24.6 \\ 
G14.631$-$00.577 & 18 19 15.21  & $-$16 30 04.50  & 25.2 \\ 
G14.991$-$00.121 & 18 18 17.32  & $-$15 58 08.30  & 46.0 \\ 
G15.094$+$00.192 & 18 17 20.82  & $-$15 43 46.50  & 25.7 \\ 
G15.607$-$00.255 & 18 19 59.34  & $-$15 29 22.80  & 65.9 \\ 
G15.665$-$00.499 & 18 20 59.75  & $-$15 33 10.00 & $-$3.0 \\ 
G16.112$-$00.303 & 18 21 09.14  & $-$15 04 00.60 & 34.5 \\ 
G16.302$-$00.196 & 18 21 07.83  & $-$14 50 54.60 & 51.8 \\ 
\hline
\end{tabular}
\end{table*}

\begin{table*}
\centering
\begin{tabular}[c]{c c c c c}
\hline
\hspace{\fill} Name (l    b) \hspace{\fill} & \hspace{\fill} RA (J200) \hspace{\fill} & \hspace{\fill} Dec (J2000) \hspace{\fill} & \hspace{\fill} V$_c$(6668)  \hspace{\fill} & Ref. \\
 (\degr \hspace{0.5cm} \degr)          & (h \hspace{0.2cm}  m \hspace{0.2cm}   s)   & (\degr \hspace{0.2cm} \arcmin \hspace{0.2cm} \arcsec) & (\kms) & \\
\hline
G16.403$-$00.181 & 18 21 16.39  & $-$14 45 09.00  & 39.2 \\ 
G16.585$-$00.051 & 18 21 09.13  & $-$14 31 48.50  & 62.1 \\ 
G16.662$-$00.331 & 18 22 19.46  & $-$14 35 39.10  & 43.0 \\ 
G16.831$+$00.079 & 18 21 09.53  & $-$14 15 08.60  & 58.7 \\ 
G16.855$+$00.641 & 18 19 09.57  & $-$13 57 57.50  & 24.2 \\ 
G16.864$-$02.159 & 18 29 24.42  & $-$15 16 04.50  & 14.9 \\ 
G16.976$-$00.005 & 18 21 44.68  & $-$14 09 48.50  & 6.5 \\ 
G17.021$-$02.403 & 18 30 36.30  & $-$15 14 28.50  & 23.5 \\ 
G17.029$-$00.071 & 18 22 05.21  & $-$14 08 51.00  & 91.3 \\ 
G17.638$+$00.157 & 18 22 26.30  & $-$13 30 12.10  & 20.7 \\ 
\\
G17.862$+$00.074 & 18 23 10.10  & $-$13 20 40.80  & 110.6\\ 
G18.073$+$00.077 & 18 23 33.98  & $-$13 09 25.00  & 55.5 \\ 
G18.159$+$00.094 & 18 23 40.18  & $-$13 04 21.00  & 58.3 \\ 
G18.262$-$00.244 & 18 25 05.70  & $-$13 08 23.20  & 75.7 \\
G18.341$+$01.768 & 18 17 58.13  & $-$12 07 24.80  & 28.0 \\ 
G18.440$+$00.045 & 18 24 23.32  & $-$12 50 52.10  & 61.8 \\ 
{\bf G18.460$-$00.004} & 18 24 36.34  & $-$12 51 08.60  & 49.4 & 2 \\ 
G18.661$+$00.034 & 18 24 51.10  & $-$12 39 22.50  & 79.0\\ 
G18.667$+$00.025 & 18 24 53.78  & $-$12 39 20.80  & 78.7\\ 
G18.733$-$00.224 & 18 25 55.53  & $-$12 42 48.90  & 45.8\\ 
\\
G18.735$-$00.227 & 18 25 56.46  & $-$12 42 50.00  & 37.9\\ 
{\bf G18.834$-$00.300} & 18 26 23.66  & $-$12 39 38.00  & 41.1  & 2 \\ 
G18.874$+$00.053 & 18 25 11.34  & $-$12 27 36.80  & 38.6 \\ 
G18.888$-$00.475 & 18 27 07.85  & $-$12 41 35.90  & 56.5 \\ 
G18.999$-$00.239 & 18 26 29.24  & $-$12 29 07.10  & 69.4 \\ 
G19.009$-$00.029 & 18 25 44.77  & $-$12 22 46.00  & 55.2 \\ 
G19.249$+$00.267 & 18 25 08.02  & $-$12 01 42.20  & 20.4 \\ 
G19.267$+$00.349 & 18 24 52.38  & $-$11 58 28.20  & 16.2 \\ 
G19.365$-$00.030 & 18 26 25.78  & $-$12 03 53.30  & 25.2 \\ 
G19.472$+$00.170 & 18 25 54.70  & $-$11 52 34.60  & 21.6 \\ 
\\
G19.496$+$00.115 & 18 26 09.16  & $-$11 52 51.70  & 121.1 \\ 
G19.609$-$00.234 & 18 27 37.99  & $-$11 56 37.60  & 40.2 \\ 
G19.612$-$00.120 & 18 27 13.48  & $-$11 53 15.70  & 53.1 \\ 
G19.612$-$00.134 & 18 27 16.52  & $-$11 53 38.20  & 56.5 \\ 
G19.614$+$00.011 & 18 26 45.24  & $-$11 49 31.40  & 32.8 \\ 
G19.667$+$00.114 & 18 26 28.97  & $-$11 43 48.90  & 14.2 \\ 
G19.701$-$00.267 & 18 27 55.52  & $-$11 52 40.30  & 43.8 \\ 
{\bf G19.755$-$00.128} & 18 27 31.66  & $-$11 45 55.00  & 123.0 & 2 \\ 
G19.884$-$00.534 & 18 29 14.37  & $-$11 50 23.00  & 46.7 \\ 
G20.081$-$00.135 & 18 28 10.32  & $-$11 28 47.60  & 43.5 \\ 
\\
G20.364$-$00.013 & 18 28 15.91  & $-$11 10 20.40  & 55.9 \\ 
G20.733$-$00.059 & 18 29 07.99  & $-$10 52 00.60  & 60.7 \\ 
G20.926$-$00.050 & 18 29 27.79  & $-$10 41 28.80  & 27.4 \\ 
G20.963$-$00.075 & 18 29 37.34  & $-$10 40 12.60  & 34.6 \\ 
G21.023$-$00.063 & 18 29 41.55  & $-$10 36 42.30  & 31.1 \\ 
G21.407$-$00.254 & 18 31 06.33  & $-$10 21 37.41  & 88.9 \\ 
G21.563$-$00.033 & 18 30 36.07  & $-$10 07 10.90  & 117.2 \\ 
G21.848$-$00.240 & 18 31 53.06  & $-$09 57 45.40  & 81.9 \\ 
G21.880$+$00.014 & 18 31 01.75  & $-$09 49 00.50  & 20.3 \\ 
G22.039$+$00.222 & 18 30 34.70  & $-$09 34 47.00  & 53.2 \\ 
\\
G22.335$-$00.155 & 18 32 29.40  & $-$09 29 29.68  & 35.6 \\ 
G22.357$+$00.066 & 18 31 44.12  & $-$09 22 12.31  & 80.1 \\ 
{\bf G22.435$-$00.169} & 18 32  43.82 & $-$09 24 33.20  & 29.5  & 1 \\ 
G23.003$+$00.124 & 18 32 44.25  & $-$08 46 10.70  & 110.5 \\ 
G23.010$-$00.411 & 18 34 40.29  & $-$09 00 38.10  & 74.7 \\ 
G23.126$+$00.395 & 18 31 59.75  & $-$08 32 09.10 & 13.8 \\ 
G23.207$-$00.377 & 18 34 55.21 & $-$08 49 14.89  & 81.6 \\
G23.257$-$00.241 & 18 34 31.26 & $-$08 42 46.70  & 63.9 \\ 
G23.365$-$00.291 & 18 34 54.13 & $-$08 38 25.60  & 82.5 \\ 
G23.389$+$00.185 & 18 33 14.32 & $-$08 23 57.47  & 75.3 \\ 
\hline
\end{tabular}
\end{table*}

\begin{table*}
\centering
\begin{tabular}[c]{c c c c c}
\hline
\hspace{\fill} Name (l    b) \hspace{\fill} & \hspace{\fill} RA (J200) \hspace{\fill} & \hspace{\fill} Dec (J2000) \hspace{\fill} & \hspace{\fill} V$_c$(6668)  \hspace{\fill} & Ref. \\
 (\degr \hspace{0.5cm} \degr)          & (h \hspace{0.2cm}  m \hspace{0.2cm}   s)   & (\degr \hspace{0.2cm} \arcmin \hspace{0.2cm} \arcsec) & (\kms) & \\
\hline
G23.437$-$00.184 & 18 34 39.25  & $-$08 31 38.50  & 102.9 \\ 
G23.440$-$00.182 & 18 34 39.18  & $-$08 31 25.40  & 96.6 \\ 
G23.484$+$00.097 & 18 33 44.05  & $-$08 21 20.60  & 87.0 \\ 
G23.657$-$00.127 & 18 34 51.56  & $-$08 18 21.30  & 82.4 \\ 
G23.707$-$00.198 & 18 35 12.36  & $-$08 17 39.36  & 76.4 \\ 
G23.818$+$00.384 & 18 33 19.50  & $-$07 55 38.10  & 76.2 \\ 
G23.885$+$00.060 & 18 34 36.84  & $-$08 01 00.70  & 45.0 \\ 
G23.901$+$00.077 & 18 34 34.92  & $-$07 59 42.20  & 35.7 \\ 
G23.966$-$00.109 & 18 35 22.21  & $-$08 01 22.47  & 70.8 \\ 
G23.986$-$00.089 & 18 35 20.09  & $-$07 59 45.00  & 65.1 \\ 
\\
G23.996$-$00.100 & 18 35 23.49  & $-$07 59 29.80  & 68.2 \\ 
G24.329$+$00.144 & 18 35 08.14  & $-$07 35 04.00  & 110.3 \\ 
G24.461$+$00.198 & 18 35 11.33  & $-$07 26 31.10  & 125.5 \\ 
G24.493$-$00.039 & 18 36 05.83  & $-$07 31 20.60  & 115.0 \\ 
G24.541$+$00.312 & 18 34 55.72  & $-$07 19 06.65  & 105.5 \\ 
G24.631$+$00.172 & 18 35 35.77  & $-$07 18 08.75  & 115.9 \\ 
G24.634$-$00.324 & 18 37 22.71  & $-$07 31 42.14  & 35.5 \\ 
G24.676$-$00.150 & 18 36 49.97  & $-$07 24 42.10  & 116.1 \\ 
G24.790$+$00.083 & 18 36 12.56  & $-$07 12 10.79  & 113.3 \\ 
G24.791$+$00.082 & 18 36 13.13  & $-$07 12 08.20  & 105.8 \\ 
\\
G24.850$+$00.087 & 18 36 18.40  & $-$07 08 51.00  & 110.0 \\ 
G24.920$+$00.088 & 18 36 25.94  & $-$07 05 07.80  & 53.3 \\ 
G24.943$+$00.074 & 18 36 31.55  & $-$07 04 16.80  & 53.2 \\ 
G25.226$+$00.288 & 18 36 16.97  & $-$06 43 18.30  & 42.0 \\ 
G25.270$-$00.434 & 18 38 56.96  & $-$07 00 49.20  & 65.9 \\ 
G25.382$-$00.182 & 18 38 15.20  & $-$06 47 56.20  & 58.2 \\ 
G25.395$+$00.034 & 18 37 30.28  & $-$06 41 17.70  & 95.4 \\ 
G25.407$-$00.170 & 18 38 15.52  & $-$06 46 16.70  & 60.8 \\ 
G25.411$+$00.105 & 18 37 16.92  & $-$06 38 30.50  & 97.2 \\ 
G25.494$+$00.062 & 18 37 35.44  & $-$06 35 13.40  & 103.8 \\ 
\\
G25.613$+$00.226 & 18 37 13.42  & $-$06 24 24.20  & 110.1 \\ 
G25.826$-$00.178 & 18 39 03.63  & $-$06 24 09.70  & 91.7 \\ 
G25.838$-$00.378 & 18 39 47.88  & $-$06 29 00.90  & $-$1.6 \\ 
G25.920$-$00.141 & 18 39 06.07  & $-$06 18 04.70  & 114.8 \\ 
G26.422$+$01.685 & 18 33 30.51  & $-$05 01 02.00  & 31.0 \\ 
G26.545$+$00.423 & 18 38 14.46  & $-$05 29 16.80  & 82.5 \\ 
G26.527$-$00.267 & 18 40 40.26  & $-$05 49 12.90  & 104.2 \\ 
G26.552$-$00.309 & 18 40 52.03  & $-$05 49 02.56  & 105.3 \\ 
G26.598$-$00.024 & 18 39 55.92  & $-$05 38 44.64 & 24.8 \\ 
G26.601$-$00.221 & 18 40 38.57  & $-$05 44 01.60 & 103.4 \\ 
\\
G26.648$+$00.018 & 18 39  52.68   & $-$05 34 54.60   & 109.4 \\ 
G27.011$-$00.039 & 18 40  44.88   & $-$05 17 09.80   & $-$18.3 \\ 
G27.221$+$00.136 & 18 40  30.54   & $-$05 01 05.39   & 118.8 \\ 
G27.286$+$00.151 & 18 40  34.51   & $-$04 57 14.40   & 34.8 \\ 
G27.365$-$00.166 & 18 41  51.06   & $-$05 01 43.50   & 99.8 \\ 
G27.500$+$00.107 & 18 41  07.38   & $-$04 47 02.30   & 87.3 \\ 
G27.757$+$00.050 & 18 41  47.99   & $-$04 34 52.60   & 99.2 \\ 
G27.783$-$00.259 & 18 42  56.96   & $-$04 41 59.00   & 98.3 \\ 
G27.784$+$00.057 & 18 41  49.58   & $-$04 33 13.80   & 111.9 \\ 
G27.869$-$00.235 & 18 43  01.55   & $-$04 36 43.10   & 20.1 \\ 
\\
G28.011$-$00.426 & 18 43  57.96   & $-$04 34  21.90   & 16.9 \\ 
G28.226$+$00.359 & 18 41  33.57   & $-$04 01  22.34   & 49.7 \\ 
G28.282$-$00.359 & 18 44  13.26   & $-$04 18  04.80   & 41.3 \\ 
G28.321$-$00.011 & 18 43  03.11   & $-$04 06  26.40   & 104.8 \\ 
G28.397$+$00.081 & 18 42  51.98   & $-$03 59  53.60   & 71.5 \\ 
G28.523$+$00.127 & 18 42  55.89   & $-$03 51  55.40   & 39.6 \\  
G28.532$+$00.129 & 18 42  56.50   & $-$03 51  21.60   & 27.0 \\  
G28.608$+$00.018 & 18 43  28.52   & $-$03 50  22.80   & 106.4 \\  
G28.687$-$00.283 & 18 44  41.54   & $-$03 54 22.10  & 92.3 \\  
G28.700$+$00.406 & 18 42  15.57   & $-$03 34 46.90  & 94.2 \\ 
\hline
\end{tabular}
\end{table*}

\begin{table*}
\centering
\begin{tabular}[c]{c c c c c}
\hline
\hspace{\fill} Name (l    b) \hspace{\fill} & \hspace{\fill} RA (J200) \hspace{\fill} & \hspace{\fill} Dec (J2000) \hspace{\fill} & \hspace{\fill} V$_c$(6668)  \hspace{\fill} & Ref. \\
 (\degr \hspace{0.5cm} \degr)          & (h \hspace{0.2cm}  m \hspace{0.2cm}   s)   & (\degr \hspace{0.2cm} \arcmin \hspace{0.2cm} \arcsec) & (\kms) & \\
\hline
{\bf G28.817$+$00.365} & 18 42 37.34  & $-$03 29 40.92  & 90.7  & 2 \\ 
G28.832$-$00.253 & 18 44 51.08  & $-$03 45 48.50  & 91.8 \\ 
G28.842$+$00.493 & 18 42 12.54  & $-$03 24 51.10  & 83.2 \\ 
G28.848$-$00.228 & 18 44 47.46  & $-$03 44 17.20  & 102.8 \\ 
G28.861$+$00.065 & 18 43 46.24  & $-$03 35 33.40  & 105.3 \\ 
G28.929$+$00.019 & 18 44 03.56  & $-$03 33 11.83  & 47.1 \\ 
G29.282$-$00.330 & 18 45 56.96  & $-$03 23 56.54  & 92.1 \\ 
G29.320$-$00.162 & 18 45 25.16  & $-$03 17 16.90  & 48.9 \\ 
G29.581$+$00.133 & 18 44 50.92  & $-$02 55 15.92  & 31.7 \\ 
G29.603$-$00.625 & 18 47 35.41  & $-$03 14 50.10  & 80.5\\ 
\\
G29.724$+$00.107 & 18 45 11.97  & $-$02 48 21.59  & 95.8 \\ 
G29.863$-$00.044 & 18 45 59.57  & $-$02 45 04.40  & 101.4 \\ 
G29.915$-$00.023 & 18 46 00.94  & $-$02 41 42.26  & 103.0 \\ 
G29.955$-$00.016 & 18 46 03.74  & $-$02 39 22.20  & 96.0 \\ 
G29.961$-$00.067 & 18 46 15.36  & $-$02 40 28.81  & 100.5 \\ 
G29.978$-$00.047 & 18 46 12.96  & $-$02 39 01.40  & 96.9 \\ 
G29.993$-$00.282 & 18 47 04.82  & $-$02 44 39.80  & 103.2 \\ 
G30.010$-$00.273 & 18 47 04.71  & $-$02 43 31.20  & 106.1 \\ 
G30.198$-$00.169 & 18 47 03.04  & $-$02 30 36.40  & 108.2 \\ 
G30.225$-$00.180 & 18 47 08.30  & $-$02 29 28.90  & 113.2 \\ 
\\ 
G30.317$+$00.070 & 18 46 25.02  & $-$02 17 40.75  & 36.1 \\ 
G30.370$+$00.482 & 18 45 02.72  & $-$02 03 33.70  & 12.4 \\ 
G30.400$-$00.296 & 18 47 52.30  & $-$02 23 16.05  & 98.2 \\ 
G30.419$-$00.232 & 18 47 40.76  & $-$02 20 30.10  & 102.9 \\ 
G30.423$+$00.466 & 18 45 12.08  & $-$02 01 13.60  & 7.5 \\ 
G30.542$+$00.011 & 18 47 02.26  & $-$02 07 17.70  & 53.1 \\ 
G30.582$-$00.141 & 18 47 39.20  & $-$02 09 19.00  & 115.5 \\ 
G30.589$-$00.043 & 18 47 18.86  & $-$02 06 17.20  & 43.0 \\ 
G30.622$+$00.082 & 18 46 55.78  & $-$02 01 07.18  & 39.6 \\ 
G30.703$-$00.068 & 18 47 36.82  & $-$02 00 53.80  & 88.2 \\ 
\\
G30.760$-$00.052 & 18 47 39.78  & $-$01 57 23.40  & 91.7 \\ 
G30.774$+$00.078 & 18 47 13.42  & $-$01 53 04.15  & 98.5 \\ 
G30.780$+$00.230 & 18 46 41.52  & $-$01 48 37.10  & 48.9 \\ 
G30.788$+$00.204 & 18 46 48.09  & $-$01 48 53.90  & 84.5 \\ 
G30.818$-$00.057 & 18 47 46.97  & $-$01 54 26.40  & 101.3 \\ 
G30.822$-$00.053 & 18 47 46.53  & $-$01 54 07.40  & 93.2 \\ 
G30.851$+$00.123 & 18 47 12.26  & $-$01 47 46.60  & 27.5 \\ 
G30.898$+$00.161 & 18 47 09.13  & $-$01 44 11.10  & 101.8 \\ 
G30.960$+$00.086 & 18 47 32.00  & $-$01 42 57.60  & 40.1 \\ 
G30.963$+$00.225 & 18 47 02.62  & $-$01 38 58.34  & 102.2 \\ 
\\
G30.972$-$00.142 & 18 48 22.07  & $-$01 48 30.30  & 77.8 \\ 
G30.973$+$00.562 & 18 45 51.69  & $-$01 29 13.30  & 19.9 \\ 
G30.980$+$00.216 & 18 47 06.47  & $-$01 38 20.00  & 111.0 \\ 
G31.047$+$00.356 & 18 46 43.85  & $-$01 30 54.15  & 81.1 \\ 
G31.059$+$00.093 & 18 47 41.35  & $-$01 37 26.20  & 16.5 \\ 
G31.076$+$00.457 & 18 46 25.44  & $-$01 26 33.50  & 25.5 \\ 
G31.122$+$00.063 & 18 47 54.68  & $-$01 34 56.90  & 48.0 \\ 
G31.158$+$00.046 & 18 48 02.40  & $-$01 33 26.80  & 41.1 \\ 
G31.182$-$00.148 & 18 48 46.41  & $-$01 37 28.10  & 46.3 \\ 
G31.253$+$00.003 & 18 48 21.92  & $-$01 29 35.68  & 41.2 \\ 
\\
G31.276$+$00.006 & 18 48 23.79  & $-$01 28 17.88  & 37.2 \\ 
G31.281$+$00.061 & 18 48 12.43  & $-$01 26 30.10  & 110.3 \\ 
G31.395$-$00.258 & 18 49 33.09  & $-$01 29 06.93  & 87.4 \\ 
G31.412$+$00.307 & 18 47 34.29  & $-$01 12 45.60  & 95.8 \\ 
G31.581$+$00.077 & 18 48 41.94  & $-$01 10 02.53  & 98.8 \\ 
G31.975$+$00.180 & 18 49 03.05  & $-$00 46 11.12  & 92.4 \\ 
G32.045$+$00.059 & 18 49 36.56  & $-$00 45 45.90  & 92.8 \\ 
G32.082$+$00.078 & 18 49 36.60  & $-$00 43 16.40  & 92.9 \\ 
G32.105$-$00.074 & 18 50 11.58  & $-$00 46 12.32  & 49.7 \\ 
G32.117$+$00.091 & 18 49 37.70  & $-$00 41 00.93  & 92.6 \\ 
\hline
\end{tabular}
\end{table*}

\begin{table*}
\centering
\begin{tabular}[c]{c c c c c}
\hline
\hspace{\fill} Name (l    b) \hspace{\fill} & \hspace{\fill} RA (J200) \hspace{\fill} & \hspace{\fill} Dec (J2000) \hspace{\fill} & \hspace{\fill} V$_c$(6668)  \hspace{\fill} & Ref. \\
 (\degr \hspace{0.5cm} \degr)          & (h \hspace{0.2cm}  m \hspace{0.2cm}   s)   & (\degr \hspace{0.2cm} \arcmin \hspace{0.2cm} \arcsec) & (\kms) & \\
\hline
G32.516$+$00.323 & 18 49 31.74  & $-$00 13 20.80  & 52.5 \\ 
G32.704$-$00.056 & 18 51 13.22  & $-$00 13 42.31  & 40.6 \\ 
{\bf G32.744$-$00.075} & 18 51 21.87  & $-$00 12 05.00  & 38.5 & 2 \\
G32.802$+$00.193 & 18 50 30.98  & $-$00 01 39.00  & 27.2 \\ 
G32.821$-$00.330 & 18 52 24.76  & $-$00 14 56.87  & 82.1 \\ 
G32.825$-$00.328 & 18 52 24.69  & $-$00 14 39.70  & 82.4 \\ 
G32.914$-$00.096 & 18 51 44.69  & $-$00 03 35.50  & 103.5 \\ 
G32.917$-$00.094 & 18 51 44.74  & $-$00 03 20.16  & 103.2 \\ 
G32.963$-$00.340 & 18 52 42.35  & $-$00 07 39.10  & 46.7 \\ 
G32.965$-$00.340 & 18 52 42.39  & $-$00 07 32.97  & 48.1 \\ 
\\
G32.992$+$00.034 & 18 51 25.58  & 00 04 08.33  & 91.9 \\ 
G33.093$-$00.073 & 18 51 59.58  & 00 06 35.50  & 103.9 \\ 
G33.133$-$00.092 & 18 52 07.82  & 00 08 12.80  & 73.2 \\
G33.199$+$00.001 & 18 51 55.34  & 00 14 19.38  & 91.2 \\ 
G33.204$-$00.010 & 18 51 58.14  & 00 14 13.61  & 91.9 \\ 
G33.317$-$00.360 & 18 53 25.30  & 00 10 43.90  & 28.1 \\ 
G33.393$+$00.010 & 18 52 14.62  & 00 24 52.90  & 105.2 \\ 
G33.424$-$00.315 & 18 53 27.40  & 00 17 40.64  & 45.6 \\ 
G33.486$+$00.040 & 18 52 18.39  & 00 30 40.20  & 121.7 \\ 
G33.641$-$00.228 & 18 53 32.56  & 00 31 39.18  & 60.3 \\ 
\\
G33.634$-$00.021 & 18 52 47.56  & 00 36 54.20  & 103.1 \\ 
G33.725$-$00.120 & 18 53 18.78  & 00 39 05.00  & 54.1 \\ 
G33.852$+$00.018 & 18 53 03.09  & 00 49 36.50  & 61.0 \\ 
G33.980$-$00.019 & 18 53 25.01  & 00 55 25.98  & 59.0 \\        
G34.096$+$00.018 & 18 53 29.94  & 01 02 39.40  & 56.1 \\ 
G34.195$-$00.593 & 18 55 51.30  & 00 51 13.58  & 61.7 \\ 
G34.244$+$00.133 & 18 53 21.44  & 01 13 44.40  & 54.9 \\ 
G34.284$+$00.184 & 18 53 15.00  & 01 17 12.99  & 51.8 \\
G34.396$+$00.222 & 18 53 19.08  & 01 24 13.80  & 55.7 \\ 
G34.411$+$00.235 & 18 53 17.99  & 01 25 25.26  & 63.1 \\ 
\\
G34.751$-$00.093 & 18 55 05.22  & 01 34 36.26  & 52.9 \\ 
G34.757$+$00.025 & 18 54 40.74  & 01 38 06.40  & 76.5 \\ 
G34.791$-$01.387 & 18 59 45.98  & 01 01 19.00  & 46.9 \\ 
G34.822$+$00.352 & 18 53 37.84  & 01 50 33.00  & 59.6 \\        
{\bf G35.132$-$00.744} & 18 58  06.14 & 01 37 07.50  & 35.4 & 2 \\ 
G35.149$+$00.809 & 18 52 35.96  & 02 20 32.03  & 75.2 \\ 
{\bf G35.197$-$00.743} & 18 58 13.05  & 01 40 35.70  & 28.5 & 2 \\  
{\bf G35.200$-$01.736} & 19 01 45.54  & 01 13 32.60  & 44.5 & 1 \\ 
G35.226$-$00.354 & 18 56 53.15  & 01 52 46.89  & 59.3 \\ 
G35.247$-$00.237 & 18 56 30.38  & 01 57 08.88  & 72.4 \\ 
\\
G35.397$+$00.025 & 18 55 50.78  & 02 12 19.10  & 89.2 \\ 
G35.417$-$00.284 & 18 56 59.02  & 02 04 55.65  & 56.0 \\ 
G35.457$-$00.179 & 18 56 40.98  & 02 09 57.16  & 55.5 \\ 
G35.588$+$00.060 & 18 56 04.22  & 02 23 28.30  & 44.1 \\ 
G35.793$-$00.175 & 18 57 16.89  & 02 27 57.91  & 60.7 \\ 
G36.115$+$00.552 & 18 55 16.79  & 03 05 05.41  & 73.1 \\ 
G36.634$-$00.203 & 18 58 55.23  & 03 12 04.72  & 77.3 \\ 
G36.705$+$00.096 & 18 57 59.12  & 03 24 06.11  & 53.0 \\ 
G36.839$-$00.022 & 18 58 39.21  & 03 28 00.90  & 61.6 \\ 
G36.918$+$00.483 & 18 56 59.78  & 03 46 03.60  & $-$35.8 \\ 
\\
G37.030$-$00.039 & 18 59 03.64  & 03 37 45.09  & 80.2 \\ 
G37.043$-$00.035 & 18 59 04.41  & 03 38 32.80  & 80.2 \\ 
G37.430$+$01.518 & 18 54 14.23  & 04 41 41.10  & 41.2 \\ 
G37.479$-$00.105 & 19 00 07.14  & 03 59 53.35  & 54.7 \\ 
G37.546$-$00.112 & 19 00 16.05  & 04 03 16.09  & 49.9 \\ 
G37.554$+$00.201 & 18 59 09.98  & 04 12 15.54  & 83.6 \\ 
G37.598$+$00.425 & 18 58 26.79  & 04 20 45.46  & 87.0 \\ 
G37.735$-$00.112 & 19 00 36.84  & 04 13 20.00  & 50.3 \\ 
G37.753$-$00.189 & 19 00 55.42  & 04 12 12.56  & 54.6 \\ 
G37.767$-$00.214 & 19 01 02.27  & 04 12 16.60  & 69.0 \\ 
\hline
\end{tabular}
\end{table*}

\begin{table*}
\centering
\begin{tabular}[c]{c c c c c}
\hline
\hspace{\fill} Name (l    b) \hspace{\fill} & \hspace{\fill} RA (J200) \hspace{\fill} & \hspace{\fill} Dec (J2000) \hspace{\fill} & \hspace{\fill} V$_c$(6668)  \hspace{\fill} & Ref. \\
 (\degr \hspace{0.5cm} \degr)          & (h \hspace{0.2cm}  m \hspace{0.2cm}   s)   & (\degr \hspace{0.2cm} \arcmin \hspace{0.2cm} \arcsec) & (\kms) & \\
\hline
G38.038$-$00.300 & 19 01 50.46  & 04 24 18.96  & 58.1 \\ 
G38.119$-$00.229 & 19 01 44.15  & 04 30 37.42  & 70.4 \\ 
G38.203$-$00.067 & 19 01 18.73  & 04 39 34.29  & 84.2 \\ 
G38.255$-$00.200 & 19 01 52.95  & 04 38 39.47  & 73.1 \\ 
G38.258$-$00.073 & 19 01 26.25  & 04 42 19.90  & 15.4 \\ 
G38.565$+$00.538 & 18 59 49.13  & 05 15 28.90  & $-$28.8 \\ 
G38.598$-$00.212 & 19 02 33.46  & 04 56 36.40  & 62.5 \\ 
G38.653$+$00.088 & 19 01 35.24  & 05 07 47.36  & $-$31.5 \\
G38.916$-$00.353 & 19 03 38.65  & 05 09 42.49  & 31.9 \\ 
G39.100$+$00.491 & 19 00 58.04  & 05 42 43.90  & 15.9 \\ 
\\
G39.388$-$00.141 & 19 03 45.31  & 05 40 42.68  & 60.2 \\ 
{\bf G40.282$-$00.219} & 19 05 41.21  & 06 26 12.69  & 73.9 & 2 \\ 
G40.597$-$00.719 & 19 08 03.29  & 06 29 12.90  & 76.2 \\ 
{\bf G40.623$-$00.138} & 19 06 01.63  & 06 46 36.50  & 31.1 & 1 \\ 
G41.075$-$00.125 & 19 06 49.04  & 07 11 06.57  & 57.5 \\ 
G41.121$-$00.107 & 19 06 50.24  & 07 14 01.49  & 36.6 \\ 
G41.123$-$00.220 & 19 07 14.85  & 07 11 00.69  & 63.4 \\ 
G41.156$-$00.201 & 19 07 14.37  & 07 13 18.10  & 56.0 \\ 
G41.226$-$00.197 & 19 07 21.37  & 07 17 08.17  & 55.4 \\ 
G41.347$-$00.136 & 19 07 21.84  & 07 25 17.27  & 11.8 \\ 
\\
G42.034$+$00.190 & 19 07 28.18  & 08 10 53.47  & 12.8 \\ 
G42.133$+$00.517 & 19 06 28.90  & 08 25 10.00  & $-$33.3 \\ 
G42.303$-$00.299 & 19 09 43.59  & 08 11 41.41  & 28.1 \\ 
G42.435$-$00.260 & 19 09 49.85  & 08 19 45.40  & 66.7 \\ 
G42.698$-$00.147 & 19 09 55.06  & 08 36 53.45  & $-$42.9 \\
G43.038$-$00.453 & 19 11 38.98  & 08 46 30.71  & 54.8 \\ 
G43.074$-$00.077 & 19 10 22.05  & 08 58 51.49  & 10.2 \\ 
G43.180$-$00.518 & 19 12 09.02  & 08 52 14.30  & 58.9 \\ 
G43.890$-$00.784 & 19 14 26.39  & 09 22 36.50  & 47.6 \\ 
G44.310$+$00.041 & 19 12 15.81  & 10 07 53.52  & 56.0 \\ 
\\
G44.644$-$00.516 & 19 14 53.76  & 10 10 07.69  & 49.5 \\
G45.071$+$00.132 & 19 13 22.12  & 10 50 53.11  & 57.7 \\ 
G45.380$-$00.594 & 19 16 34.14  & 10 47 01.60  & 53.3 \\ 
G45.445$+$00.069 & 19 14 18.31  & 11 08 59.40  & 50.0 \\
G45.473$+$00.134 & 19 14 07.36  & 11 12 16.00  & 65.7 \\ 
G45.493$+$00.126 & 19 14 11.35  & 11 13 06.20  & 57.2 \\ 
G45.804$-$00.356 & 19 16 31.08  & 11 16 12.01  & 59.9 \\ 
G46.066$+$00.220 & 19 14 56.07  & 11 46 12.98  & 23.5 \\ 
G46.115$+$00.387 & 19 14 25.52  & 11 53 25.99  & 58.2 \\ 
G48.902$-$00.273 & 19 22 10.33  & 14 02 43.51  & 71.8 \\ 
\\
G49.043$-$01.079 & 19 25 22.25  & 13 47 19.50  & 36.6 \\ 
G49.265$+$00.311 & 19 20 44.85  & 14 38 26.91  & $-$4.7 \\ 
G49.349$+$00.413 & 19 20 32.44  & 14 45 45.44  & 67.9 \\
G49.416$+$00.326 & 19 20 59.21  & 14 46 49.60  & $-$12.1 \\ 
G49.417$+$00.324 & 19 20 59.82  & 14 46 49.10  & $-$26.6 \\ 
G49.470$-$00.371 & 19 23 37.90  & 14 29 59.30  & 63.8 \\ 
G49.471$-$00.369 & 19 23 37.60  & 14 30 05.40  & 73.5 \\ 
G49.482$-$00.402 & 19 23 46.19  & 14 29 47.10  & 51.8 \\ 
G49.489$-$00.369 & 19 23 39.82  & 14 31 04.90  & 56.2 \\        
G49.599$-$00.249 & 19 23 26.61  & 14 40 16.99  & 63.0 \\ 
\\
G49.617$-$00.360 & 19 23 52.81  & 14 38 03.30  & 50.4 \\ 
G50.035$+$00.582 & 19 21 15.45  & 15 26 49.20  & $-$5.1 \\ 
G50.315$+$00.676 & 19 21 27.47  & 15 44 18.60  & 30.1 \\ 
G50.779$+$00.152 & 19 24 17.41  & 15 54 01.60  & 49.1 \\ 
G51.679$+$00.719 & 19 23 58.87  & 16 57 41.80  & 7.3 \\ 
G51.818$+$01.250 & 19 22 17.95  & 17 20 06.50  & 46.5 \\
G52.199$+$00.723 & 19 24 59.84  & 17 25 17.90  & 3.7 \\ 
G52.663$-$01.092 & 19 32 36.07  & 16 57 38.40  & 65.8 \\ 
G52.922$+$00.414 & 19 27 34.96  & 17 54 38.14  & 39.1 \\ 
G53.036$+$00.113 & 19 28 55.49  & 17 52 03.11  & 10.0 \\
\hline
\end{tabular}
\end{table*}

\begin{table*}
\centering
\begin{tabular}[c]{c c c c c}
\hline
\hspace{\fill} Name (l    b) \hspace{\fill} & \hspace{\fill} RA (J200) \hspace{\fill} & \hspace{\fill} Dec (J2000) \hspace{\fill} & \hspace{\fill} V$_c$(6668)  \hspace{\fill} & Ref. \\
 (\degr \hspace{0.5cm} \degr)          & (h \hspace{0.2cm}  m \hspace{0.2cm}   s)   & (\degr \hspace{0.2cm} \arcmin \hspace{0.2cm} \arcsec) & (\kms) & \\
\hline
G53.142$+$00.071 & 19 29 17.58  & 17 56 23.21  & 24.6 \\ 
G53.618$+$00.036 & 19 30 23.01  & 18 20 26.68  & 18.9 \\ 
G56.963$-$00.235 & 19 38 17.10  & 21 08 05.40  & 29.9 \\ 
G57.610$+$00.025 & 19 38 40.74  & 21 49 32.70  & 38.9 \\ 
G58.775$+$00.644 & 19 38 49.13  & 23 08 40.20  & 33.3 \\ 
G59.634$-$00.192 & 19 43 50.00  & 23 28 38.80  & 29.6 \\ 
G59.783$+$00.065 & 19 43 11.25  & 23 44 03.30  & 27.0 \\ 
G59.833$+$00.672 & 19 40 59.33  & 24 04 46.50  & 38.0 \\ 
G60.575$+$00.186 & 19 45 52.48  & 24 17 42.99  & 3.4 \\ 
G70.181$+$01.741 & 20 00 54.16  & 33 31 30.88  & $-$26.8 \\      
\\
G71.522$+$00.385 & 20 12 57.91  & 33 30 26.95  & 8.2 \\ 
G73.063$+$01.796 & 20 08 10.20  & 35 59 23.70  & 5.9 \\ 
G75.782$+$00.342 & 20 21 44.05  & 37 26 36.91  & $-$1.0 \\ 
G78.122$+$03.633 & 20 14 26.04  & 41 13 33.39  & $-$7.7 \\ 
G78.886$+$00.708 & 20 29 24.94  & 40 11 19.28  & $-$6.9 \\ 
G79.736$+$00.991 & 20 30 50.67  & 41 02 27.60  & $-$5.5 \\ 
{\bf G81.722$+$00.571} & 20 39 01.05  & 42 22 49.18  &$-$2.7  & 3 \\
G81.744$+$00.590 & 20 39 00.38  & 42 24 36.91  & 4.0 \\ 
G81.752$+$00.590 & 20 39 01.99  & 42 24 59.08  & $-$5.7 \\ 
G81.871$+$00.780 & 20 38 36.42  & 42 37 34.56  & 6.3 \\
\\
G94.602$-$01.796 & 21 39 58.26  & 50 14 20.96  & $-$40.8 \\
G97.521$+$03.172 & 21 32 13.00  & 55 52 56.00  & $-$71.2 \\
G107.288$+$05.638 & 22 21 22.50 & 63 51 13.00  & $-$8.5 \\
G108.184$+$05.519 & 22 28 51.40 & 64 13 41.31  & $-$11.0\\
G108.766$-$00.986 & 22 58 51.18 & 58 45 14.37  & $-$46.3 \\
{\bf G109.871$+$02.114} & 22 56 17.90  & 62 01 49.65  & $-$3.7  &  3 \\
G111.255$-$00.769 & 23 16 10.33 & 59 55 28.43  & $-$38.9 \\
G121.298$+$00.659 & 00 36 47.35 & 63 29 02.16  & $-$23.3 \\
G123.066$-$06.309 & 00 52 24.19 & 56 33 43.17  & $-$29.4 \\
G136.845$+$01.167 & 02 49 33.59 & 60 48 27.95  & $-$45.0 \\
\\
G173.482$+$02.446 & 05 39 13.05  & 35 45 51.29  & $-$13.0 \\
G173.698$+$02.886 & 05 41 37.40  & 35 48 49.00  & $-$23.8 \\
G174.201$-$00.071 & 05 30 48.01  & 33 47 54.61  & 1.5 \\
G188.793$+$01.030 & 06 09 06.96  & 21 50 41.23  & $-$5.3 \\
G188.946$+$00.886 & 06 08 53.34  & 21 38 29.16  & 10.8 \\
G189.471$-$01.216 & 06 02 08.37  & 20 09 20.10  & 18.8 \\
G189.777$+$00.344 & 06 08 35.30  & 20 39 06.59  & 4.6 \\
G192.600$-$00.048 & 06 12 54.02  & 17 59 23.32  & 4.6 \\
G196.454$-$01.677 & 06 14 37.05  & 13 49 36.16  & 15.1 \\
G206.543$-$16.355 & 05 41 44.15  & $-$01 54 44.90  & 12.1 \\
\\
G209.016$-$19.398 & 05 35 13.95   & $-$05  24 09.40  & $-$1.5 \\
G212.063$-$00.741 & 06 47 12.90   & 00  26 07.00  & 43.3 \\
G213.705$-$12.597 & 06 07 47.86   & $-$06  22 56.52  & 10.7 \\
G232.620$+$00.995 & 07 32 09.78   & $-$16  58 12.57 & 22.7 \\
\hline
\end{tabular}
\tablebib{(1) \citet{caswellvaile1995};
(2) \citet{avison2016}; (3) \citet{baudry1997}
}
\end{table*}

\begin{table*}
\centering
\caption{6035 and 6031\,MHz OH Zeeman pair candidates identified in the present sample. The peak velocities ($V_{\mathrm{f}}$) and peak flux densities ($S_{\mathrm{f}}$) obtained by fitting Gaussian components to the spectra of LHC and RHC features are listed. The demagnetized velocities ($V_{\mathrm{d}}$), magnetic field strength (B) and estimates of the magnetic field strength from the literature (B$_{\mathrm{lit}}$) are also given. The reliability parameter describes two cases of Zeeman splitting estimation: both polarised spectral features showed single Gaussian components (A) or were blended (B).}\label{tab:zeeman}
\begin{tabular}{lcc@{\hskip -5pt}ccccccl}
\hline
         & \multicolumn{2}{c}{LHC} & & \multicolumn{2}{c}{RHC} & & & & \\
         \cline{2-3} \cline{5-6}
    Name & $V_{\mathrm{f}}$ & $S_{\mathrm{f}}$  && $V_{\mathrm{f}}$ & $S_{\mathrm{f}}$ & $V_{\mathrm{d}}$ &  B & Reliability &B$_{\mathrm{lit}}$\\
     & (km~s$^{-1}$) & (Jy) && (km~s$^{-1}$) & (Jy) & (km~s$^{-1}$) & (mG) & & (mG) \\
\hline
G11.034$+$00.062 &  23.93 & 1.83 && 23.71 & 0.92 & 23.82 & $-$3.9 & B &$-$7.7$^1$; $-6.1^4$ \\ 
G11.904$-$00.141 &  42.76 & 1.60 && 42.90 & 1.00 & 42.83 & $+$2.5 & A &$<$0.5$^1$; $+1.6^4$\\
G12.209$-$00.102 &  16.96 & 1.77 && 17.34 & 2.00 & 17.15 & $+$6.8 & B &\\
G12.681$-$00.182 &  59.72 & 1.93 && 60.39 & 1.57 & 60.06 & $+$12.0 & B &\\ 
G15.035$-$00.677 &  21.43 & 38.29 && 21.46 & 29.11 & 21.45 & $+$0.6 & A &$<$0.5$^1$;$+$1.5$^2$;$+0.9^4$\\
                 &  22.63 & 19.55 && 22.63 & 21.63 & 22.63 & $-$0.1 & A & $-0.2^4$\\ 
                 &  23.60 & 7.53 && 23.36 & 3.88 & 23.48 & $-$4.2 & A & $-5.4^4$\\
                 &  21.53 & 6.40 && 21.54 & 2.85 & 21.54$^a$ & $+$0.2 & A &$+0.5^4$\\
G25.650$+$01.049 &  39.53 & 2.81 && 39.25 & 1.77 & 39.34 & $-$5.4 & A &\\
G28.201$-$00.049 &  94.54 & 3.05 && 94.94 & 3.45 & 94.74 & $+$7.1 & B &$+$9.0$^2$;$+$6.2$^3$;$+$7.5$^3$\\
                 &  94.47 & 0.75 && 95.02 & 1.31 & 94.74$^a$ &$+$7.0 & B &\\
G35.025$+$00.350 &  45.27 & 1.40 && 45.57 & 3.23 & 45.42 & $+$5.4 & B &$+$5.0$^1$\\
G43.149$+$00.013 &  11.15 & 3.22 && 10.84 & 3.03 & 11.00 & $-$5.5 & A&$-$4.3$^2$\\
G43.796$-$00.127 &  40.18 & 0.56 && 40.44 & 0.59 & 40.31 & $+$4.6 & A &$+$3.6$^1$\\
G45.467$+$00.053 &  65.03 & 4.53 && 64.88 & 17.02 & 64.95 & $-$2.8 & A &$-$3.2$^3$\\ 
                 &  66.36 & 12.43 && 66.29& 6.35 & 66.32 & $-$1.2 & A&\\
                 &  67.71 & 4.11 && 68.05 & 3.31 & 67.88 & $+$6.0 & A &\\
                 &  65.15 & 1.76 && 64.84 & 1.58 & 65.00$^a$ & $-$3.9 & A &\\
                 &  66.38 & 2.26 && 66.25 & 0.97 & 66.31$^a$ & $-$1.6 & A &\\
G48.990$-$00.299 &  67.52 & 1.74 && 67.64 & 3.91 & 67.58 & $+$2.1 & A &\\
G49.490$-$00.388 &  52.26 & 1.72 && 52.55 & 1.75 & 52.40 & $+$5.0 & B &$+$5.0$^2$, $+$3.9$^3$ \\
                 &  54.75 & 9.05 && 54.97 & 10.09 & 54.86 & $+$3.8 & B &\\
                 &  56.51 & 3.97 && 56.77 & 7.41 & 56.64 & $+$4.8 & A &\\
                 &  57.65 & 2.21 && 57.94 & 1.99 & 57.79 & $+$5.3 & A &\\
                 &  62.92 & 1.55 && 63.08 & 2.40 & 63.00 & $+$2.9 & A &\\
                 &  52.30 & 4.70 && 52.67 & 2.92 & 52.48$^a$ & $+$4.8 & A &\\
G69.540$-$00.976 &  0.02 & 2.97 && $-$0.55 & 2.68 & $-$0.27 & $-$10.1 & B &$-$2.7$^3$\\
                 &  14.52 & 7.60 && 14.48 & 13.41 & 14.50 & $-$0.9 & B &\\
                 &  14.09 & 1.77 && 13.79 & 2.62 & 13.94$^a$ & $-$3.8 & A &$-$3.2$^3$;$-$4.2$^3$\\
G80.861$+$00.383 &  $-$11.02 & 2.30 && $-$11.18 & 1.59 & $-$11.10 & $-$2.8 & A &$-$4.3$^3$\\ 
                &  $-$1.23 & 1.29 && $-$1.08 & 2.00 & $-$1.15 & $+$1.6 & A &\\
                 &  $-$11.02 & 1.34 && $-$11.25 & 1.52 &$-$11.13$^a$ & $-$2.9 & A& $-$3.2$^3$;$-$2.7$^3$\\
G81.871$+$00.781 &  7.44 & 8.63 && 7.85 & 13.96 & 7.65 & $+$7.4 & B &$+$7.8$^3$\\
                 &  8.67 & 4.97 && 9.13 & 1.47 & 8.90 & $+$8.3 & B &$+$7.5$^3$\\
G85.410$+$00.003 &  $-$32.90 & 0.66 && $-$32.97 & 1.98 & $-$32.93 & $-$1.2 & A &\\
G98.036$+$01.446 &  $-$62.77 & 0.58 && $-$62.58 & 0.58 & $-$62.68 & $+$3.4 & B &\\
                 &  $-$61.47 & 1.81 && $-$61.41 & 1.65 & $-$61.44 & $+$1.1 &A &\\
G108.766$-$00.986 &  $-$44.69 & 1.05 && $-$45.32 & 3.89 & $-$45.00 & $-$11.2 & B & \\
                  &  $-$44.18 & 1.11 && $-$44.82 & 1.96 & $-$44.50 & $-$11.4 & B & \\
G111.542$+$00.777 &  $-$59.53 & 1.37 && $-$59.50 & 1.11 & $-$59.51 & $+$0.6 & B &\\
G133.947$+$01.064 &  $-$43.33 & 180.0 && $-$43.00 & 125.19 & $-$43.17 & $+$5.9 & B &\\ 
                  &  $-$43.43 & 34.57 && $-$42.78 & 81.54 & $-$43.10$^a$ & $+$8.2 & B & \\
                  &  $-$42.82 & 19.97 && $-$42.11 & 20.25 & $-$42.47$^a$ & $+$9.0 &B &\\
\hline
\end{tabular}
\tablefoot{$^a$ 6031\,MHz transition. {\bf References.} $^1$ \citet{caswellvaile1995}, $^2$ \citet{caswell2003}, $^3$ \citet{baudry1997}, $^4$ \citet{green2015}}
\end{table*}

\end{appendix}

\end{document}